\documentclass[%
reprint,
secnumarabic,
nobibnotes,
 amsmath,amssymb,
  aps,
prfluids,
lettersize,
onecolumn
]{revtex4-2}

\usepackage{epstopdf, epsfig}   

\usepackage{graphicx}
\usepackage{dcolumn}
\usepackage{bm}
\usepackage{hyperref}

\newcommand\calD{\mathcal{D}} 

\def\refcite#1{Ref.~\cite{#1}}
\def\reffig#1{Fig.~\ref{fig:#1}}

\def\refsec#1{\S\ref{sec:#1}}
\def\refeq#1{Eq.~(\ref{eq:#1})}

\def\utau {{u_\tau}}

\def\adj {{\dagger}}
\def\calD{\mathcal{D}}
\def\Vmu{V_{M\bfu}}
\def\Vmt{V_{M\theta}}
\def\Vtu{V_{T\bfu}}
\def\Vtt{V_{T\theta}}

\def\state {\mathbf{q}}
\def\vel   {\mathbf{u}}
\def\forc  {\mathbf{f}}

\def\statehat {\widehat{\state}}
\def\velhat   {\widehat{\vel}}
\def\temphat  {\widehat{\theta}}
\def\forchat  {\widehat{\forc}}

\def\bfu   {\mathbf{u}}
\def\bff   {\mathbf{f}}

\def\bfA   {\mathbf{A}}
\def\bfB   {\mathbf{B}}
\def\bfC   {\mathbf{C}}

\def\bfP   {\mathbf{P}}

\def\bfI   {\mathbf{I}}

\def\bfX   {\mathbf{X}}
\def\bfU   {\mathbf{U}}

\def\Pr{\mbox{{Pr}}}
\def\Rey{\mbox{{Re}}}
\newcommand\Ra{\mbox{{Ra}}}
\newcommand\Ri{\mbox{{Ri}}}
\newcommand\Nu{\mbox{{Nu}}}
\newcommand\Ritau{\Ri_\tau}
\newcommand\Retau{\Rey_\tau}

\newcommand\Reb{\Rey_b}

\begin{document}


\title{Non-normal energy amplifications in stratified turbulent channels}

\author{Carlo Cossu}
 \email{Carlo.Cossu @ CNRS.Fr}
 \affiliation{Hydrodynamics, Energetics and Atmospheric Environment Research Laboratory (LHEEA)\\
 CNRS - Nantes Universit\'e - \'Ecole Centrale Nantes, F-44000 Nantes, France}

 \date{\today}

\begin{abstract}
The influence of stable and unstable stratification on the amplification of coherent structures in turbulent channel flows is investigated by computing the linear response to stochastic forcing near the turbulent mean flow. 
The velocity and thermal responses to momentum and thermal forcing are considered separately. 
It is found that, consistently with results of previous direct numerical simulations, the influence  of the mean flow stratification on stochastic forcing amplifications is non-negligible only for streamwise-elongated large-scale structures.
Unstable stratification is found to enhance the peak variance of the response, except for the velocity response to thermal forcing, and to increase the spanwise wavelength of the most amplified structures.
Stable stratification induces opposite effects.
The different spanwise wavelengths maximizing the different types of variance amplifications, all converge to approximately six channels half-widths when approaching the linear instability threshold where large-scale coherent rolls become linearly unstable.
We show that in the presence of even moderately unstable stratification, the  profiles of turbulent buoyancy and momentum fluxes and of $rms$  vertical velocity of all types of most amplified stochastic responses are nearly indistinguishable from those of the critical mode becoming unstable at the critical Richardson number. 
For all considered stratification levels, the two most energetic POD modes are found to contribute to more than 90\% of the variance of the response, except for the thermal response to thermal forcing.
We conclude that the same mechanism underlies the onset of the instability of coherent large-scale rolls at the critical Richardson number and the amplification of coherent large-scale structures at subcritical Richardson numbers.  
The process leading to the onset of the instability of large-scale rolls is therefore gradual  and the increasing response variance associated to increasingly unstable mean flow stratification as well as the increase of the optimal spanwise wavelength of the most amplified mechanically forced streaks, can be both interpreted as precursors of the linear instability of large-scale rolls. 

\end{abstract}

\maketitle


\section{Introduction \label{sec:intro}}

We are interested in the influence of mean flow stratification on the amplification of coherent structures in wall-bounded shear flows.
While a clear theoretical understanding has been reached in the case of laminar flows by means of stability analyses predicting optimal non-modal amplifications and the onset of linear modal instabilities, such is not the case for turbulent flows where a clear theoretical  understanding of the genesis and the main characteristics of large-scale coherent motions is still lacking.
Such an understanding would be beneficial to many applications, ranging from the design of heat-exchangers to weather forecasting and climate sensitivity analyses where better models of large-scale coherent structures are sought. 
In this study we choose to focus on the sole interactions of buoyancy and shear by considering the Poiseuille-Rayleigh-B\'enard flow, thus removing additional effects such as e.g. ground roughness, Coriolis acceleration, three-dimensionality of the mean velocity profiles which would be encountered in geophysical applications.
In this configuration, the viscous, thermally-conducting fluid is confined in a plane channel between two horizontal isothermal walls enforcing either a destabilizing stratification (when the ground is hotter than the top wall) or a stabilizing one (in the opposite case) and is driven by a (streamwise) pressure gradient.

The modal stability of laminar steady solutions of the Poiseuille-Rayleigh-B\'enard flow is well understood.
In the absence of stratification the laminar Poiseuille solution is known to become linearly unstable to Tollmien-Schlichting waves when the Reynolds number exceeds the critical value $\Rey_c=5772$ \cite{Orszag1971,Drazin1981}.
This critical Reynolds number does not change when destabilizing stratifications are enforced \cite{Gage1968}.
Excessive destabilizing stratifications, however, induce the linear instability of Rayleigh-B\'enard convection rolls with spanwise wavelength $\lambda \approx 4h$ (where $h$ is the channel half-width) when the Rayleigh number exceeds the critical value $\Ra_c=1708$ \cite{Pellew1940,Drazin1981}.  
The value of the critical Rayleigh number, initially determined in the absence of mean flow (no pressure gradient, $\Rey=0$),  remains unchanged for non-zero Reynolds numbers \cite{Gage1968} with rolls aligning with the streamwise direction of the Poiseuille flow.

The determination of the critical Reynolds and Rayleigh numbers and the associated neutral modes, however, is not sufficient to fully characterize the dynamics of the considered Poiseuille-Rayleigh-Bénard flow which can become turbulent even for Reynolds numbers significantly lower than the critical $\Rey_c$ \cite{Schmid2001}.
This subcritical transition has been related to the potential of linearly stable laminar base flows to sustain very large amplifications of small perturbations exploiting the highly non-normal nature of the linearized Navier-Stokes operator. 
This potential has been investigated by computing the largest energy amplifications of initial conditions and forcing and the corresponding optimal inputs and outputs.
In plane channels,  the most amplified perturbations are streamwise streaks, i.e. streamwise-elongated spanwise-periodic low- and high-speed regions, which are optimally induced by streamwise vortices  with most amplified spanwise wavelengths $\lambda_y \approx 3h$ \cite{Gustavsson1991,Butler1992,Farrell1993b,Reddy1993,Trefethen1993}.  
Stabilizing stratifications  are found to reduce the optimal energy amplifications \cite{Biau2004} while destabilizing stratifications do increase them \cite{Jerome2012}. 
In the latter case, streamwise-uniform perturbations remain the most amplified ones with the most amplified spanwise wavelength gradually drifting from $\lambda \approx 3h$ in the unstratified case to $\lambda \approx 4h$ when approaching the critical Rayleigh number while the amplification of small wavelengths is left substantially unaffected by stratification  \cite{Jerome2012}. 

When the Reynolds number and/or the Rayleigh number are sufficiently large, the channel flow is turbulent and is characterized by persistent large-scale coherent structures such as large-scale streaks \cite{Jimenez1998,Jimenez2004,Hutchins2007} and convection rolls \cite{Garai2014,Pirozzoli2017} containing a substantial fraction of the turbulent kinetic energy.
The resemblance of these coherent structures to their laminar counterparts has motivated linear stability analyses of turbulent mean flows.
In a first approach, the `quasi-laminar' one, nonlinear fluctuations are considered as a forcing to the Navier-Stokes operator linearized near the turbulent mean flow whose selective amplification of turbulent fluctuations is then analyzed \cite{Butler1993,Farrell1993b,McKeon2010,McKeon2017}.
In the present study we follow  a different approach based on the triple decomposition of the flow fields into temporal mean, coherent part of the fluctuations  and residual random fluctuations \cite{Reynolds1972}.
In this approach, which has been adopted in a large number of previous linear analyses of turbulent mean flows \cite{Bottaro2006,delAlamo2006,Cossu2009,Crouch2009,Pujals2009,Hwang2010, Hwang2010c,Willis2010,Tammisola2016,Pickering2021}, the amplification of \textit{coherent} fluctuations 
is studied based on linear operators which embed the effects of the turbulent stresses induced by the random part of the fluctuations.
In this context, the turbulent mean flow in plane channels is found to be linearly stable in the absence of stratification  \cite{Reynolds1967,Reynolds1972} and  becomes linearly unstable for sufficiently large destabilizing stratifications, where a critical mode consisting of coherent streamwise-uniform large-scale rolls of  spanwise wavelength $\lambda_y \approx 6\,h$ becomes unstable at the critical friction Richardson number $\Ri_{\tau,c}=-0.86$  \cite{Cossu2022}.
In the linearly stable regime, however, the turbulent mean flow is still able to sustain non-normal energy amplifications despite the additional damping associated to turbulent diffusion.
Most previous research has considered unstratified channels where streamwise streaks are found to be the most amplified structures emerging in response to an initial condition or to harmonic or stochastic forcing  \cite{delAlamo2006,Pujals2009,Hwang2010c}.
In the unstratified case, optimally-amplified logarithmic layer streaks are found to be almost-self-similar geometrically and their amplification scales with the spanwise wavenumber $k_y$ as $k_y^{-\gamma}$  with $\gamma=2$ when considering the optimal response to harmonic forcing, $\gamma=1$ for the variance of the response to stochastic forcing, and $\gamma=0$ when considering optimal temporal amplifications of initial conditions \cite{Hwang2010c,Willis2010,Cossu2017}.
The amplifications of buffer-layer and large-scale streaks, however, depart from the logarithmic-layer algebraic scaling with peak amplifications respectively found near $\lambda^+ \approx 90$ (scaling in wall units) and $\lambda \approx 3.5-5h$ (scaling in the outer length scale $h$) \cite{Hwang2010c}, consistently with the size of the most energetic turbulent structures.

The influence of stratification on the non-normal amplification of coherent structures in turbulent flows has been addressed only recently.
Ahmed \textit{et al.} \cite{Ahmed2021} have investigated the low-rank properties of the resolvent operator in $\Retau=180$ stably stratified turbulent channels by means of the quasi-laminar formulation.  
Zasko \textit{et al.} \cite{Zasko2022} explored higher $\Retau=O(1000)$ Reynolds numbers in the turbulent Couette flow by including eddy viscosity and thermal diffusivity in the linear operator and finding an increase of optimal temporal energy amplifications for stabilizing stratifications.
Madhusudanan \textit{et al.}  \cite{Madhusudanan2022} and Cossu \cite{Cossu2022} have considered the effect of destabilizing temperature gradients including eddy viscosity and thermal diffusivity in the linear operator, the former computing the response to impulsive forcing, the latter the critical Rayleigh and Richardson numbers for the onset of the linear instability of large-scale convection rolls.

The effect of an unstable stratification on non-normal energy amplifications, however, has not been investigated yet in turbulent channels nor in other turbulent canonical flows thus leaving unanswered a number of significant questions:
Do energy amplifications increase with destabilizing stratification?
If yes, to what extent?
Which coherent perturbations are the most influenced by stratification?
What are the main features of the most amplified coherent perturbations? 
Are optimally amplified coherent structures a precursor of the critical mode that is destabilized for sufficiently large unstable stratification or do they have distinct characteristics?

The goal of this study is to answer the questions raised above by computing the coherent response to stochastic forcing in stratified turbulent channels and evaluating the mean amplification of the forcing.
The effect of stable stratification will also be investigated because only the Couette flow was previously considered \cite{Zasko2022} with a non-quasi-laminar approach.
To gain a clear view of the underlying amplification mechanisms, the amplifications of  velocity and temperature coherent fluctuations will be computed separately in response to momentum and heating stochastic forcing, departing from the customary use of a compound norm \cite{Jerome2012,Ahmed2021,Zasko2022}.
The paper is organized as follows: The mathematical formulation of the problem is introduced in \refsec{bground},
the results are presented in \refsec{Results} and discussed in \refsec{concl} where some conclusions are drawn.
 Additional details are provided in the appendix.

\section{Background
\label{sec:bground}}

\subsection{Flow configuration and linear model for coherent structures
\label{sec:mathset}}

We consider the pressure-driven flow in a plane channel delimited by two horizontal walls located at $z=\pm 1$ orthogonal to the gravitational field $-g\,\mathbf{e}_z $, where we denote by $x$, $y$ and $z$ the streamwise, spanwise and vertical coordinates made dimensionless with respect to the channel half-width $h$ and by  $\mathbf{e}_z$ the vertical unit vector. 
The fluid, whose thermal expansion coefficient is $\beta$, is viscous and thermally conducting with kinematic viscosity $\nu$ and thermal diffusivity $\alpha$.
For the considered turbulent flows it is customary to express the distance from the walls also in wall units as $z^+=(h - |z|) u_\tau / \nu$, where $u_\tau=\sqrt{|\tau_w|/\rho}$ is the characteristic velocity associated to the wall shear stress $\tau_w$.
A constant temperature difference $\Delta \Theta=\Theta(z=1)-\Theta(z=-1)$ is maintained between the two walls, which are assumed to be isothermal, resulting in a vertical heat flux $Q$.  
Note that, following the usual  convention, positive (negative) values of $\Delta \Theta$  and $Q$  correspond to stabilizing (destabilizing) mean temperature gradients.

We use a linear Newtonian eddy closure to model small-amplitude coherent velocity $\vel=(u,v,w)$, pressure $p$ and temperature  $\theta$ fluctuations 
to the time-averaged mean flow $\bfU=U(z) \mathbf{e}_x$, $P(z)$ and $\Theta(z)$. 
This model has been used in a number of previous studies  \cite{Reynolds1972,Pujals2009,Hwang2010,Hwang2010c,Willis2010,Moarref2012,Illingworth2018,Morra2019,Madhusudanan2019} and has been extended to include buoyancy effects under the Boussinesq approximation \cite{Madhusudanan2022,Cossu2022,Zasko2022}:
\begin{eqnarray}
\label{eq:LinNS}
\partial_t \vel   &=& - \nabla \vel \cdot \bfU-\nabla \bfU \cdot \vel
 + \Ritau\, \theta\, \mathbf{e}_z
 -\nabla p + \nabla \cdot \left[ \nu_T \left(\nabla \vel +\nabla \vel^T \right) \right] + \bff_\bfu, \\
\label{eq:LinTH}
\partial_t \theta &=& - \nabla \theta \cdot \bfU-\nabla \Theta \cdot \vel
    +\nabla \cdot \left( \alpha_T \nabla \theta \right) + f_\theta,
\end{eqnarray}
where $\bff_\bfu$ and $f_\theta$ are the momentum and thermal forcing terms and the equations are made dimensionless in terms of the channel half-width $h$, the temperature difference $\Delta \Theta$ and the friction velocity $\utau$.
The system depends explicitly on the friction Richardson number 
$\Ritau= h \beta g \Delta \Theta / \utau^2$ (positive in the stably stratified case and negative in the unstably stratified case) and on the friction Reynolds number  $\Retau=h \utau /\nu$ and the Prandtl number $\Pr=\nu/\alpha$ via the effective kinematic viscosity $\nu_T$, the effective thermal diffusivity $\alpha_T$ and the associated mean flow profiles.
For the sake of comparison with previous investigations,  results will be discussed also in terms of the Rayleigh number 
$\Ra=(2h)^3 g \beta \Delta \Theta / (\alpha \nu)$ and of the bulk Reynolds number  $\Reb=2 h U_b / \nu$  based on the mean bulk velocity $U_b$.

\begin{figure}
\centerline{
    \includegraphics[width=0.45\textwidth]{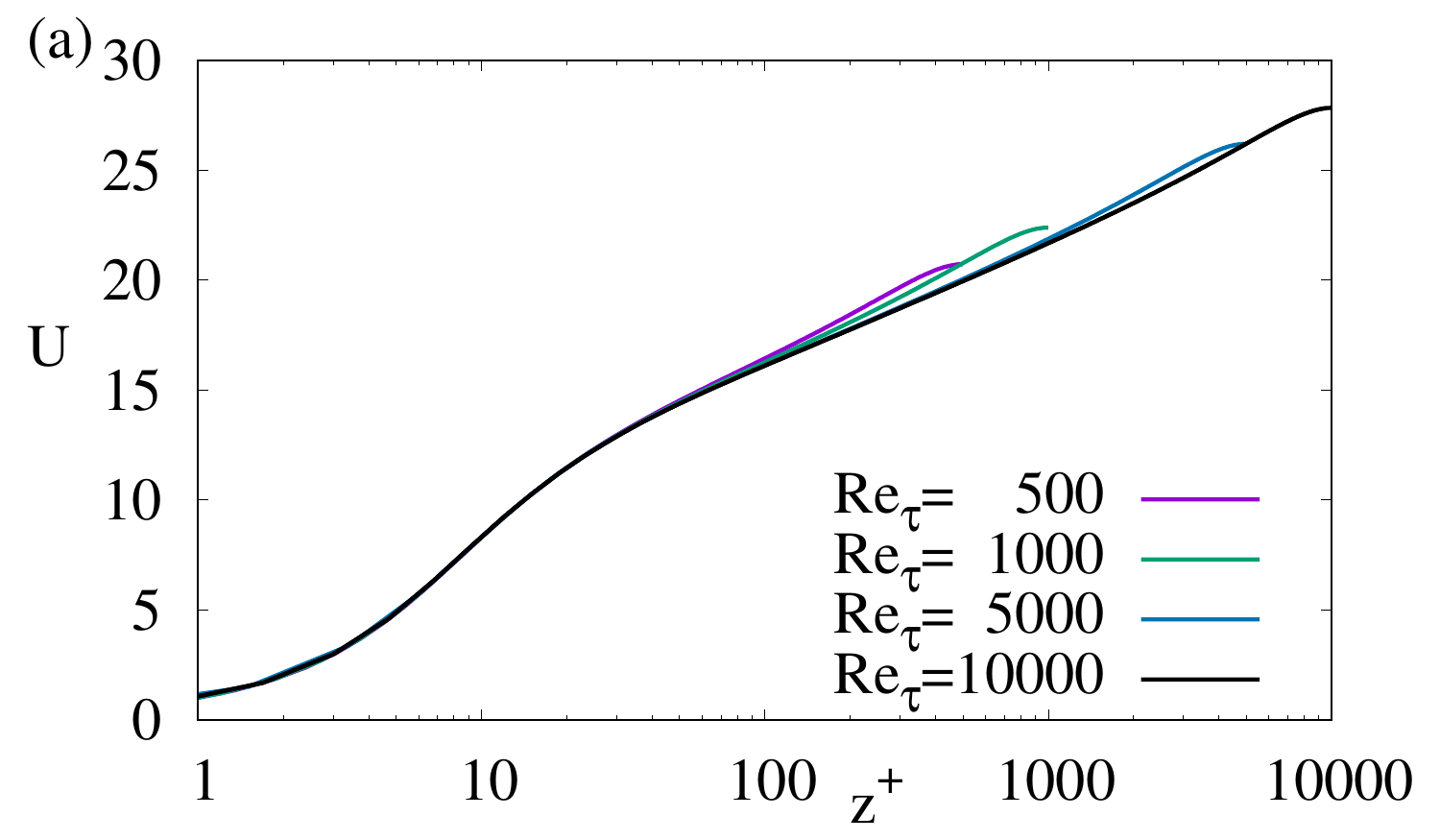}      
    \includegraphics[width=0.45\textwidth]{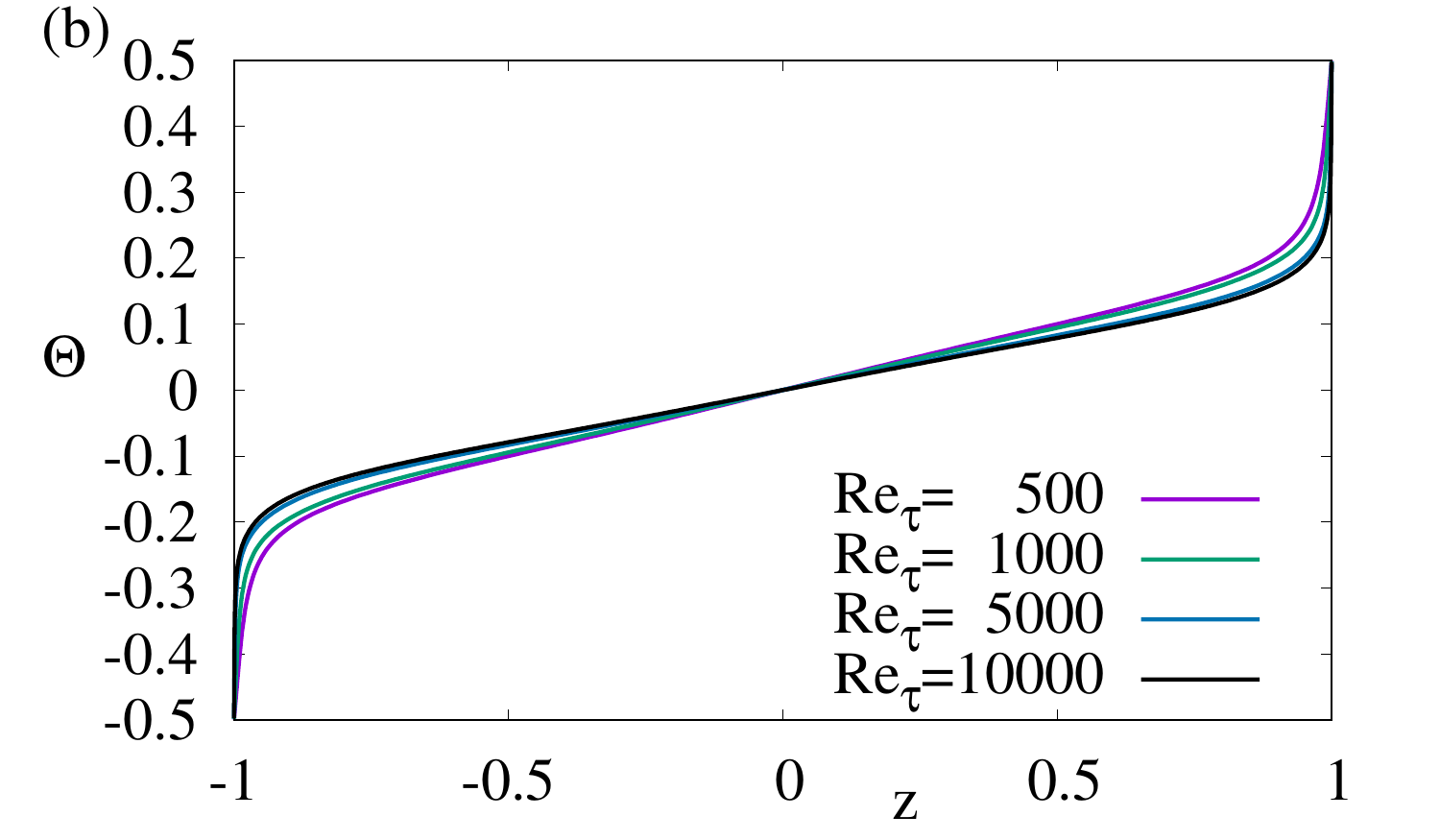}    
}
  \caption{
Vertical profiles of the temporally-averaged mean streamwise velocity $U$ (expressed in wall units, panel $a$) and mean temperature $\Theta$ (normalized with respect to $\Delta \Theta$, panel $b$) for selected friction Reynolds numbers $\Retau$.
} 
\label{fig:MeanFlow}
\end{figure}

Fourier transforms in the horizontal coordinates and standard manipulations are used to reduce the system given by equations~(\ref{eq:LinNS}-\ref{eq:LinTH}) to the following system for the wall-normal velocity wall-normal vorticity and the temperature Fourier modes  ${\widehat{w}}(z,t)$, ${\widehat{\zeta}}(z,t)$, ${\temphat}(z,t)$ of streamwise and spanwise wavenumbers $k_x$ and $k_y$ forming the state vector $\statehat$:
\begin{eqnarray}
\label{eq:OSSQTH}
\partial_t \statehat = \bfA \statehat + \bfB \forchat;~~~~
\bfA=\left[\begin{array}{ccc}
   \Delta^{-1}\mathcal{L_{OS}} & 0 & \Ritau\, k^2  \Delta^{-1} \\
   -i\,k_y U' & \mathcal{L_{SQ}} & 0 \\
   -\Theta' & 0 & \mathcal{L_{\theta}} 
\end{array}\right],~~
\bfB = \left[\begin{array}{cccc}
   -i k_x \Delta^{-1}\calD & - k^2 \Delta^{-1} & -i k_y\Delta^{-1} \calD & 0\\
   i k_y & 0 & -i k_x & 0\\
   0 & 0 & 0 & 1
\end{array}\right],
\end{eqnarray}
where 
$\statehat=\left\{\widehat{w}, \widehat{\zeta},  \temphat \right\}^T$,
$\forchat=\left\{\widehat{f}_u, \widehat{f}_v, \widehat{f}_w, \widehat{f}_\theta \right\}^T$ and
the generalized Orr-Sommerfeld,  Squire  and $\mathcal{L_{\theta}}$ linear operators, including the effects of eddy viscosity and eddy thermal diffusivity, are defined as:
\begin{eqnarray}
\label{eq:OS}
\mathcal{L_{OS}}&=&-ik_x(U \Delta-U'')
       +\nu_T \Delta^2+2 \nu_T' \Delta \calD +\nu_T''(\calD^2+k^2),\\
\mathcal{L_{SQ}}&=&-ik_x U + \nu_T \Delta+\nu_T' \calD,\\
\mathcal{L_{\theta}}&=&-ik_x U + \alpha_T \Delta+\alpha_T' \calD
\label{eq:SQTH}
\end{eqnarray}
with $\calD$ and $'$ denoting $d/d z$,  $k^2=k_x^2+k_y^2$ and $\Delta=\calD^2-k^2$.
No-slip and isothermal boundary conditions, are enforced on both walls: 
$\widehat{w}(\pm 1)=0$, 
$\calD\widehat{w}(\pm 1)=0$, 
$\widehat{\zeta}(\pm 1)=0$,
$\temphat(\pm 1)=0$.
The mean flow velocity $U(z)$ and temperature $\Theta(z)$ profiles, a sample of which is shown in \reffig{MeanFlow}, and  the associated $\nu_T(z)$ and  $\alpha_T$ profiles appearing in equations~(\ref{eq:OSSQTH}), (\ref{eq:OS}) and (\ref{eq:SQTH}) are based on the extended Cess's model described in Appendix \ref{app:ExtCess}, which has been widely used in linear analyses of unstratified channels \cite{Reynolds1967,Waleffe1993,Farrell1996b,delAlamo2006,Pujals2009,Hwang2010c,Morra2019} and has recently been extended to the stratified case \cite{Cossu2022}. 

\subsection{Response to stochastic forcing
\label{sec:StoFo}}

Coherent perturbations to the turbulent mean flow are linearly stable as long as the friction Richardson does not exceed the critical value $\Ri_{\tau,c}=-0.86$ found in Ref.~\cite{Cossu2022} by means of the modal stability analysis of the linear operator $\bfA$.
In the linearly stable regime it is of interest to quantify the linear system response to stochastic forcing representing the effect of neglected nonlinear terms.
We therefore follow previous investigations \cite{Farrell1993,Farrell1993b,Bamieh2001,Jovanovic2005,Hwang2010,Hwang2010c}
in considering a zero-mean
($\langle{\forchat}\rangle=\mathbf{0}$) stochastic forcing  with 
$\langle\forchat(t)\forchat^H(t')\rangle = \bfP \delta(t-t')$ where $\langle \cdot \rangle$ denotes the ensemble average and the usual choice $\bfP=\bfI$ is made.
This forcing induces a stochastic response with covariance 
$\langle \statehat \statehat^H \rangle = \bfX$
which, as $t \rightarrow \infty$, tends to the solution of the algebraic Lyapunov equation \cite{Farrell1993b}:
\begin{equation}\label{eq:Lyapunov}
\mathbf{A\mathbf{}X}+\mathbf{X}\bfA^\adj+\mathbf{BPB}^\adj=0,
\end{equation}
where the superscript $^\adj$ denotes adjoint operators.
Because of their different physical significance, it is important to distinguish the momentum forcing 
$\widehat{\bff}_\bfu=\{\widehat{f}_u, \widehat{f}_v, \widehat{f}_w,\}^T$ 
from the thermal forcing 
$\widehat{f}_\theta$. 
To this end, we separately compute 
the solution $\bfX_M$ of the Lyapunov equation when only the mechanical forcing is active (i.e. $\widehat{f}_\theta=0$), having forcing covariance $\bfP_M$, and
the solution $\bfX_T$ of the Lyapunov equation where only the thermal forcing is active (i.e. $\widehat{\bff}_\bfu = \mathbf{0}$) with forcing covariance $\bfP_T$, where $\bfP_M+\bfP_T=\bfI$ and  
\begin{eqnarray}
\label{eq:Pforc}
\bfP_M = \left[\begin{array}{cccc}
   1 & 0 & 0 & 0 \\
   0 & 1 & 0 & 0\\
   0 & 0 & 1 & 0\\
   0 & 0 & 0 & 0
\end{array}\right] ~~~;~~~
\bfP_T = \left[\begin{array}{cccc}
   0 & 0 & 0 & 0 \\
   0 & 0 & 0 & 0\\
   0 & 0 & 0 & 0\\
   0 & 0 & 0 & 1
\end{array}\right].
\end{eqnarray}
As the velocity and temperature can be retrieved from the state vector as 
$\velhat=\bfC_\bfu \statehat$ and 
$\temphat=\bfC_\theta \statehat$, 
the velocity and temperature covariance are given by 
$\langle \velhat \velhat^H \rangle =\bfC_\bfu \bfX \bfC^\adj_\bfu$,
$\langle \temphat \temphat^H \rangle =\bfC_\theta \bfX \bfC^\adj_\theta$, where $\bfX$ is either $\bfX_M$ or $\bfX_T$ and 
\begin{eqnarray}
\bfC_\bfu = \frac{1}{k^2} \left[\begin{array}{ccc}
   i\alpha \calD &  -i\beta & 0 \\
   k^2 &        0     & 0 \\
    i\beta\calD & i\alpha & 0\\
\end{array}\right] ;~~~~
\bfC_\theta = \left[\begin{array}{ccc}
   0 & 0 & 1 
\end{array}\right].
\end{eqnarray}
The following four ratios will be used to quantify the respective amplification of the variance of momentum and thermal forcing into velocity and temperature coherent perturbations variance: 
\begin{eqnarray}
\label{eq:VMamp}
\Vmu &=&  \frac{Tr \langle \velhat \velhat^H \rangle}{Tr \langle \forchat_\bfu \forchat_\bfu^H \rangle}
,~~~~~
\Vmt =  U_e^2 \frac{ Tr \langle \temphat \temphat^H \rangle}{Tr \langle \forchat_\bfu \forchat_\bfu^H \rangle}
\\
\Vtu &=& \frac{1}{U_e^2} \frac{Tr \langle \velhat \velhat^H \rangle}{Tr \langle \widehat{f}_\theta \widehat{f}_\theta^H \rangle}
,~~~~~
\Vtt = \frac{Tr \langle \temphat \temphat^H \rangle}{Tr \langle \widehat{f}_\theta \widehat{f}_\theta^H \rangle}.
\label{eq:VTamp}
\end{eqnarray}
where, e.g. $Tr \langle \velhat \velhat^H\rangle=\int_{-1}^1 \left(\langle \hat u^* \hat u\rangle +\langle \hat v^* \hat v\rangle +\langle \hat w^* \hat w\rangle \right)\,dz$ is the velocity variance.
Note that in (\ref{eq:VMamp}) and (\ref{eq:VTamp}) the velocity and momentum forcing variances are normalized with respect to  the square of the centreline mean velocity $U_e$ corresponding to the maximum  velocity variation $\Delta U$ of the mean flow to make them comparable to temperature and heating variances which are normalized by $\Delta \Theta$.

To quantify the level of coherence in the response to stochastic forcing and to identify the most relevant coherent structures we also compute the eigenvalues and eigenfunctions of $\langle \velhat \velhat^H \rangle$ and $\langle \temphat \temphat^H \rangle$.
Covariance operators being Hermitian, their eigenvalues $\sigma_j$ are real and correspond to a set of mutually orthogonal eigenfunctions often referred to as proper orthogonal decomposition (POD) modes, `empirical orthogonal functions'  or Karhunen-Lo\`eve modes. 
As the sum of the $\sigma_j$ eigenvalues is equal to the total variance of the response $V=\sum_j \sigma_j$, the ratio $\sigma_j / \sum_j \sigma_j$ represents the contribution of the $j$-th mode to the response variance; the corresponding eigenfunction provides the associated coherent structure emerging in the response. 

Standard methods are used to numerically compute variance amplifications.
The system reported in \refeq{OSSQTH}) and the $\bfC_\bfu$, $\bfC_\theta$ operators 
are discretized in the vertical direction by means of a Chebyshev-collocation method using the discretized differentiation operators of \refcite{Weideman2000} which embed the appropriate boundary conditions.
The stochastic response is obtained by solving \refeq{Lyapunov} with the \texttt{lyap} function in \texttt{matlab}. 
The codes have been derived from those used and validated in Refs.~\cite{Hwang2010,Hwang2010c,Cossu2022}. 
The results in the present study are obtained by using a number of collocation points ranging from 129 to 513 depending on the Reynolds number, as in Refs.~\cite{Pujals2009,Hwang2010c,Cossu2022}.

\section{Results
\label{sec:Results}}

\subsection{Effects of stratification on stochastic forcing amplifications at $\mathbf{Re_\tau}=1000$
\label{sec:Re1000}}

\begin{figure}
\centerline{
 \includegraphics[width=0.28\textwidth]{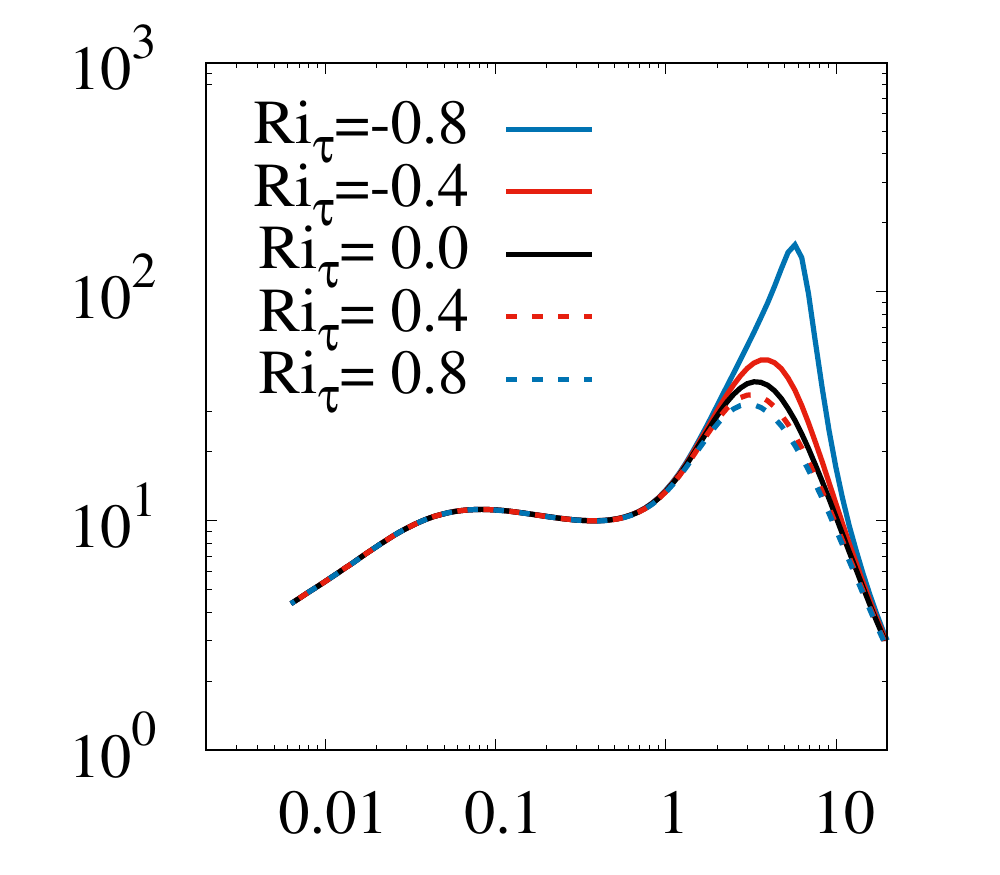} \hspace{-4mm}  
 \put(-130,120){{$(a)$}}
 \put(-130,95){{$\Vmu$}}
 \put(-52,1){{$\lambda_y$}}
 \includegraphics[width=0.28\textwidth]{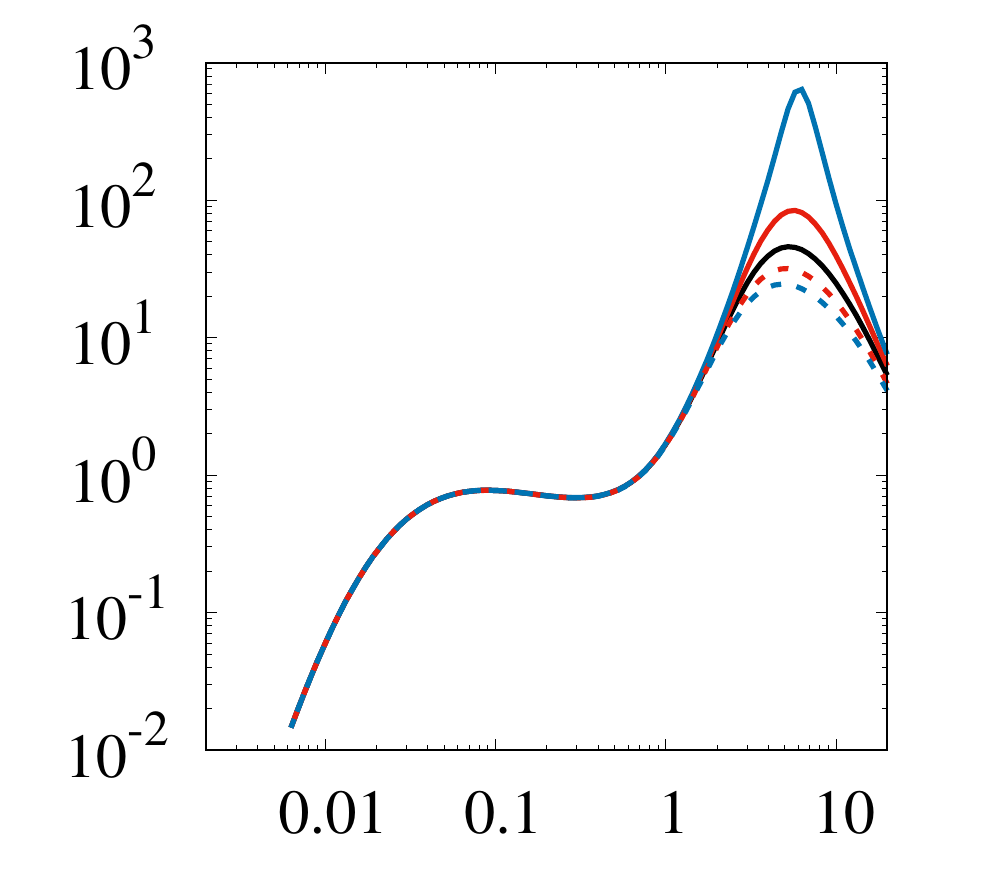} \hspace{-4mm}
 \put(-130,120){{$(b)$}}
 \put(-132,85){{$\Vmt$}}
 \put(-52,1){{$\lambda_y$}}
 \includegraphics[width=0.28\textwidth]{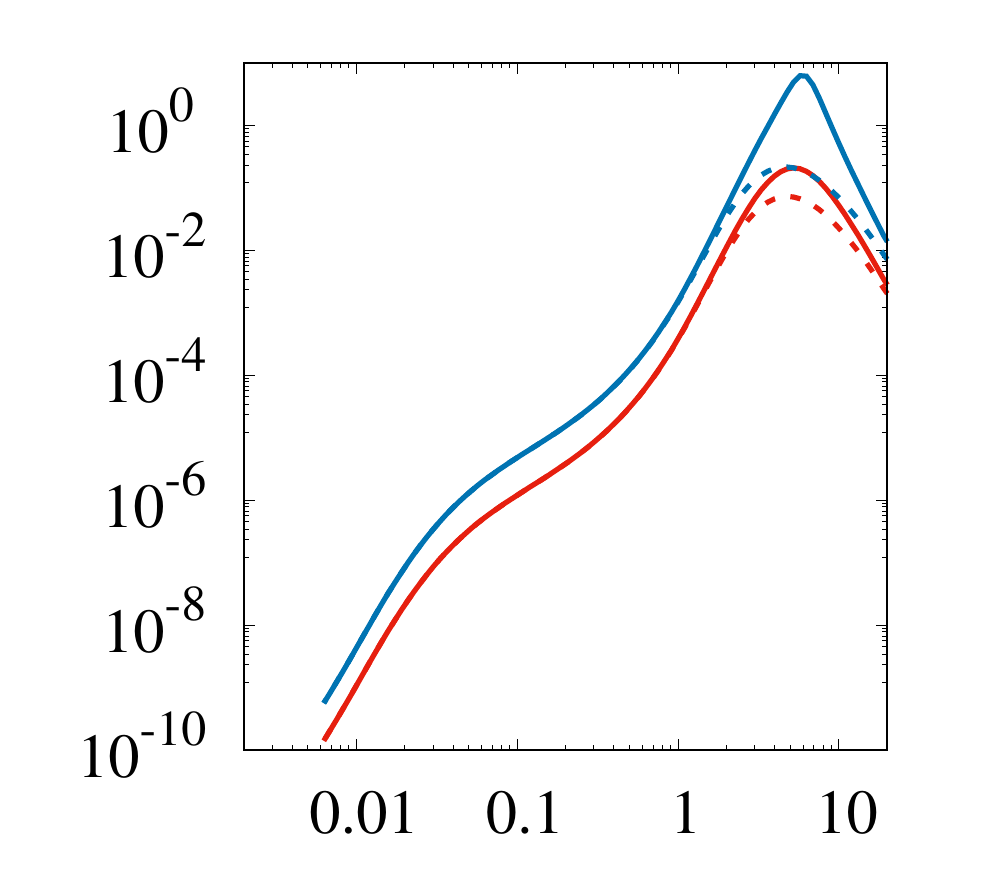} \hspace{-4mm}
 \put(-125,120){{$(c)$}}
 \put(-130,80){{$\Vtu$}}
 \put(-52,1){{$\lambda_y$}}
 \includegraphics[width=0.28\textwidth]{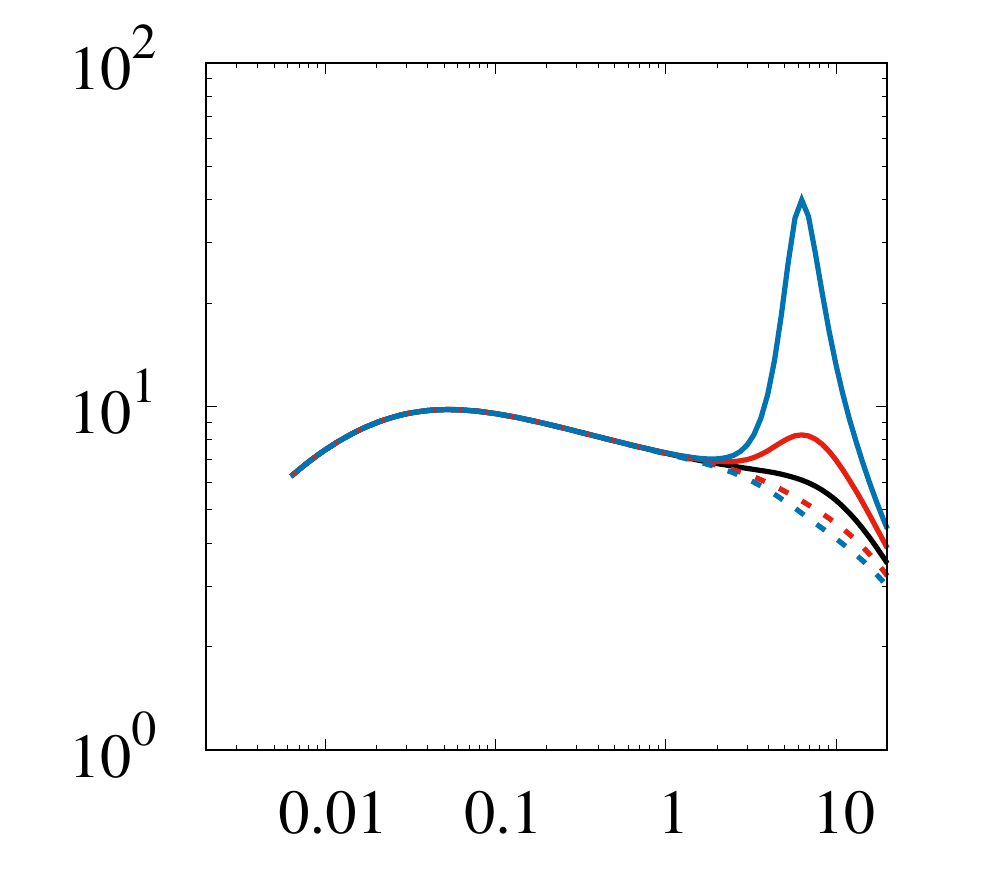} \hspace{-4mm}
 \put(-130,120){{$(d)$}}
 \put(-125,84){{$\Vtt$}}
 \put(-52,1){{$\lambda_y$}}
}
\vspace*{-1mm}
\centerline{
 \includegraphics[width=0.28\textwidth]{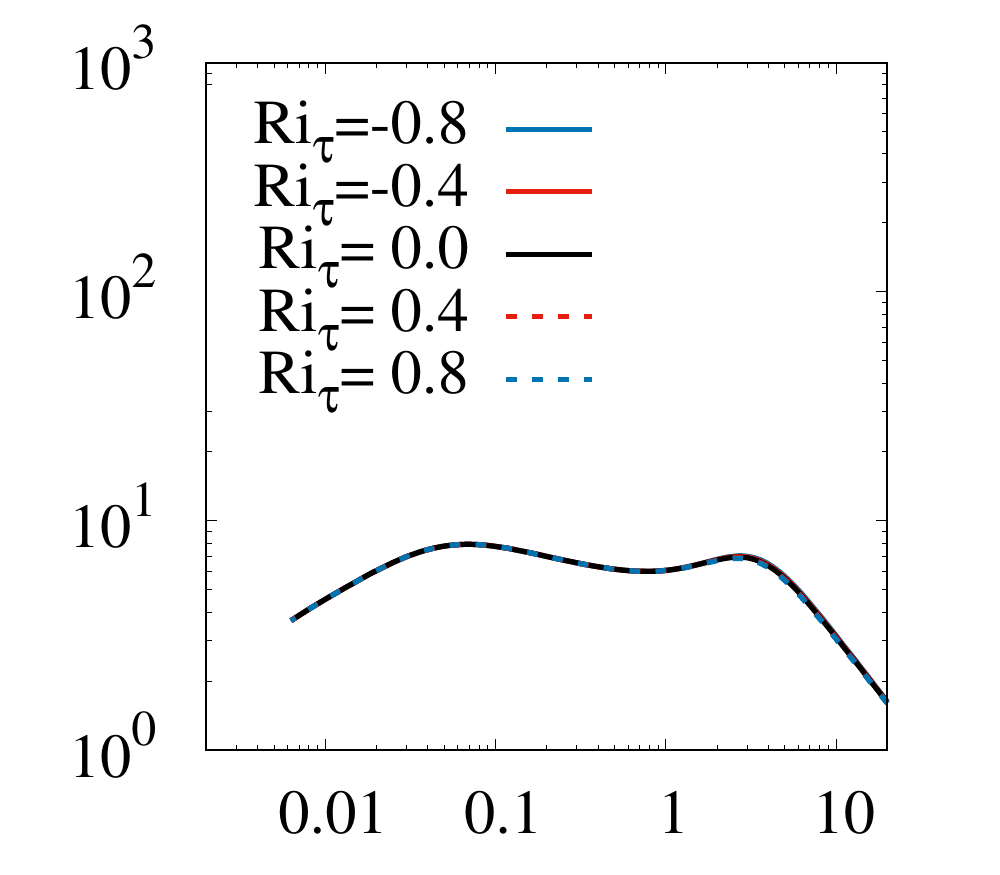} \hspace{-4mm}
 \put(-130,120){{$(e)$}}
 \put(-130,95){{$\Vmu$}}
 \put(-52,1){{$\lambda_y$}}
 \includegraphics[width=0.28\textwidth]{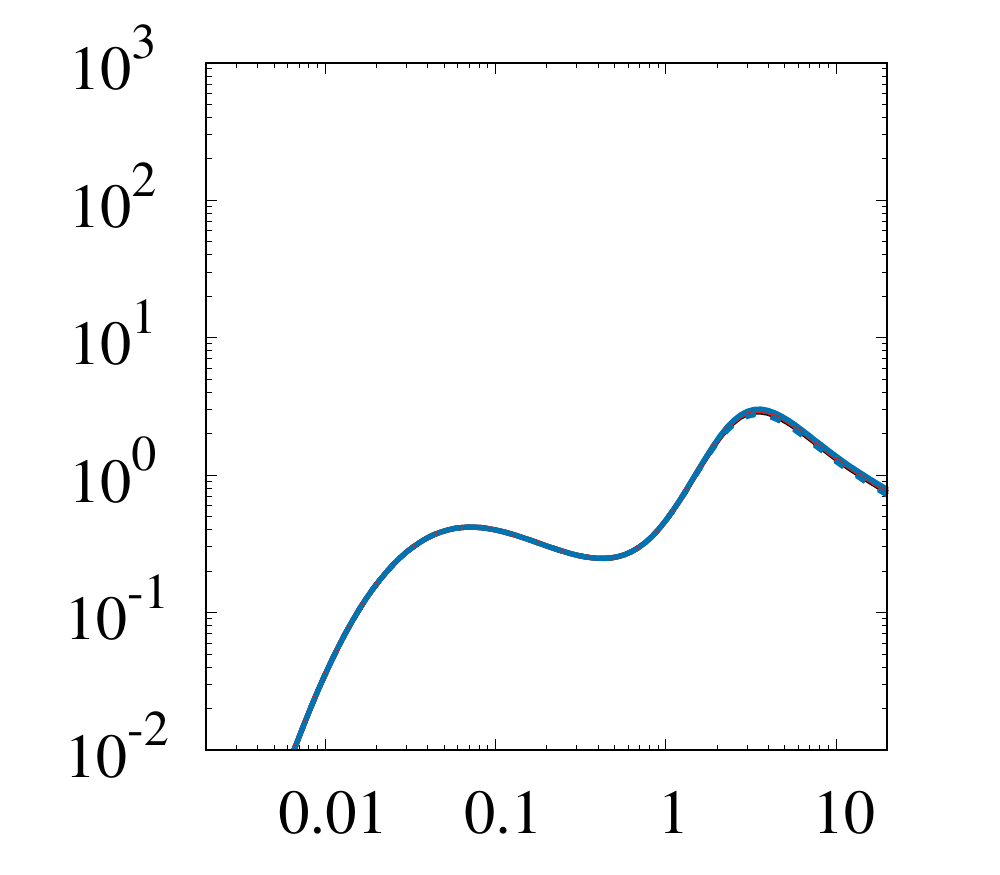} \hspace{-4mm}
 \put(-130,120){{$(f)$}}
 \put(-132,85){{$\Vmt$}}
 \put(-52,1){{$\lambda_y$}}
 \includegraphics[width=0.28\textwidth]{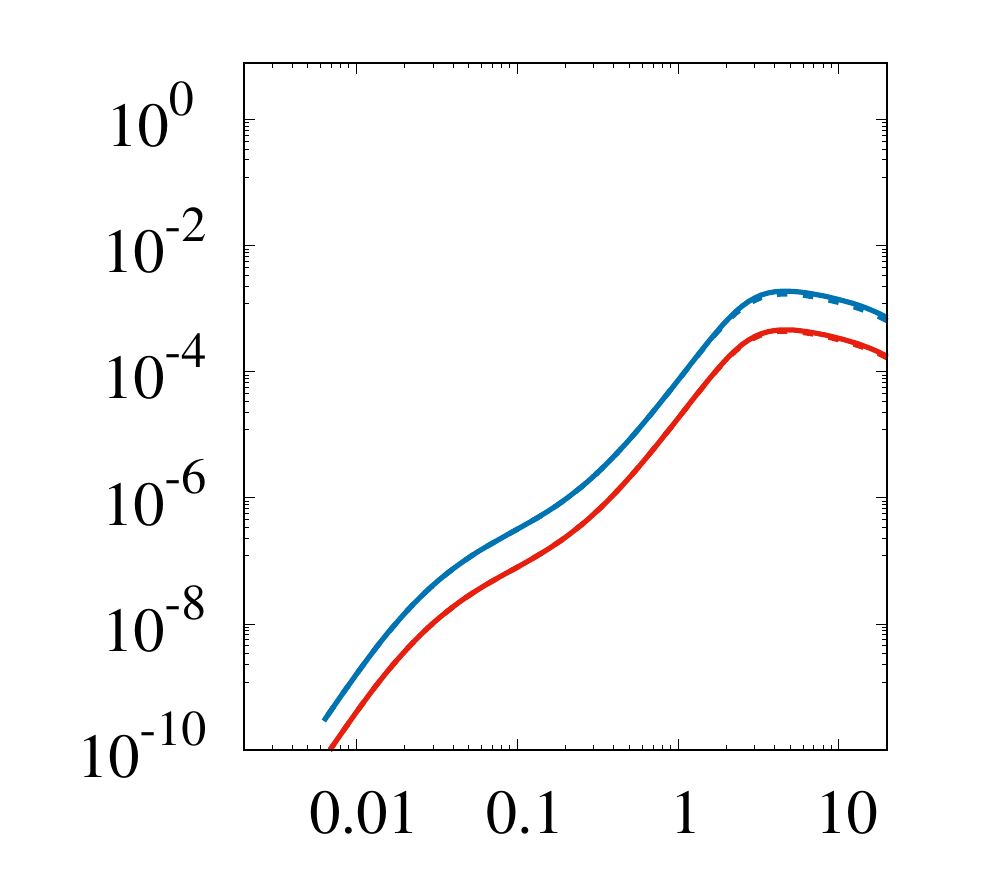} \hspace{-4mm}
 \put(-125,120){{$(g)$}}
 \put(-130,80){{$\Vtu$}}
 \put(-52,1){{$\lambda_y$}}
 \includegraphics[width=0.28\textwidth]{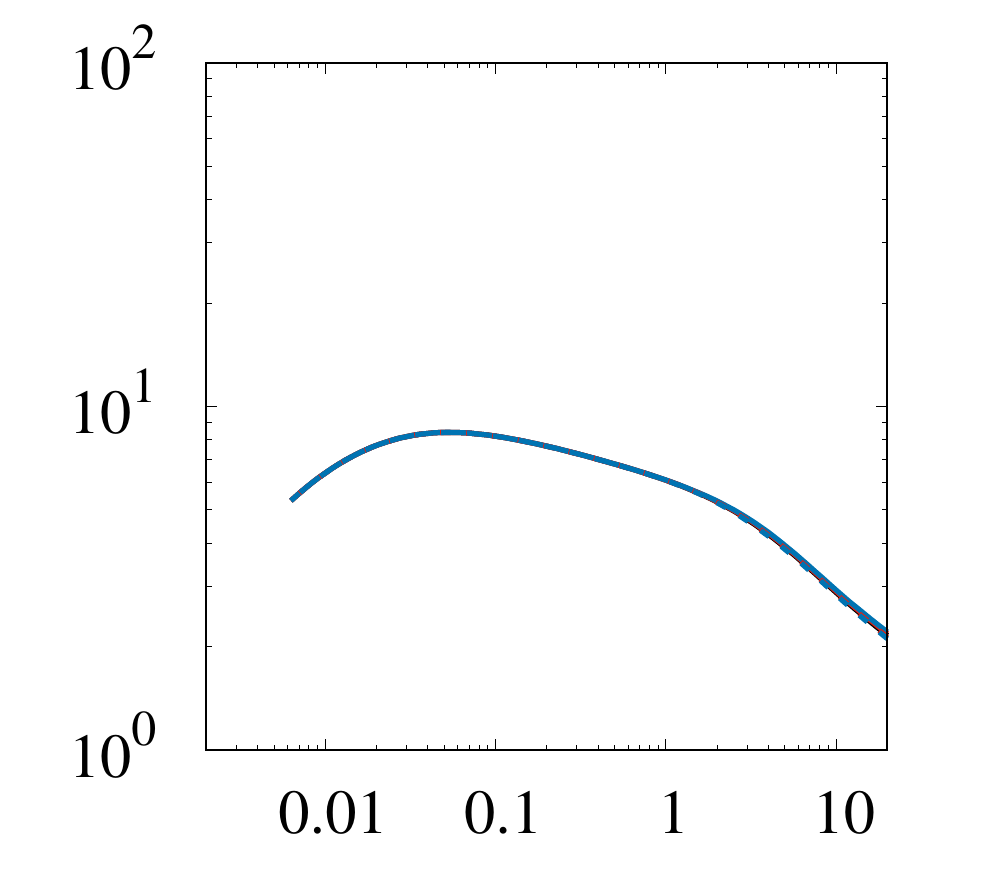} \hspace{-4mm}
 \put(-130,120){{$(h)$}}
 \put(-125,84){{$\Vtt$}}
 \put(-52,1){{$\lambda_y$}}
}
\vspace*{-3mm}
\caption{Premultiplied variance amplification ratios
$\Vmu$   (panels $a$ and $e$),
$\Vmt$ (panels $b$ and $f$),
$\Vtu$   (panels $c$ and $g$) and
$\Vtt$ (panels $d$ and $h$) reported as a function of the spanwise wavelength $\lambda_y=2 \pi / k_y$ for 
$\Retau=1000$ and selected values of $\Ritau$. 
The amplifications of streamwise-uniform perturbations (with $k_x=0$) are reported in the top row (panels $a$-$d$) and those of perturbations with $\lambda_x = 2 \lambda_y$ in the bottom row (panels $e$-$h$).
} 
\label{fig:VPRElambdaChiR1000}
\end{figure}

In this section we investigate the influence of stratification on variance amplifications for the (fixed) friction Reynolds number $\Retau=1000$. Responses to stochastic forcing are computed for Richardson numbers ranging from $\Ritau=0.8$ (in the stably stratified regime) to $\Ritau=-0.8$ (in the unstably stratified regime).
The Prandtl number is set to $\Pr=1$ for all the results presented in this paper.
The variance amplification ratios, premultiplied by the spanwise wavenumber $k_y$, 
are reported in \reffig{VPRElambdaChiR1000} as a function of the spanwise wavelength $\lambda_y=2 \pi / k_y$ for selected values of $\Ritau$.
Two types of perturbations are considered: streamwise-uniform ($k_x=0$) perturbations  (top row of  \reffig{VPRElambdaChiR1000}),  which are the most amplified ones, and perturbations with $\lambda_x = 2 \lambda_y$ (bottom row of  \reffig{VPRElambdaChiR1000}) corresponding to wavelengths typical of the self-sustained process \cite{Hwang2010b,Hwang2011,Cossu2017}. 
Additional results pertaining to intermediate values of the streamwise wavelength are reported in Appendix~B. 

\begin{figure}
\centerline{
 \includegraphics[width=0.44\textwidth]{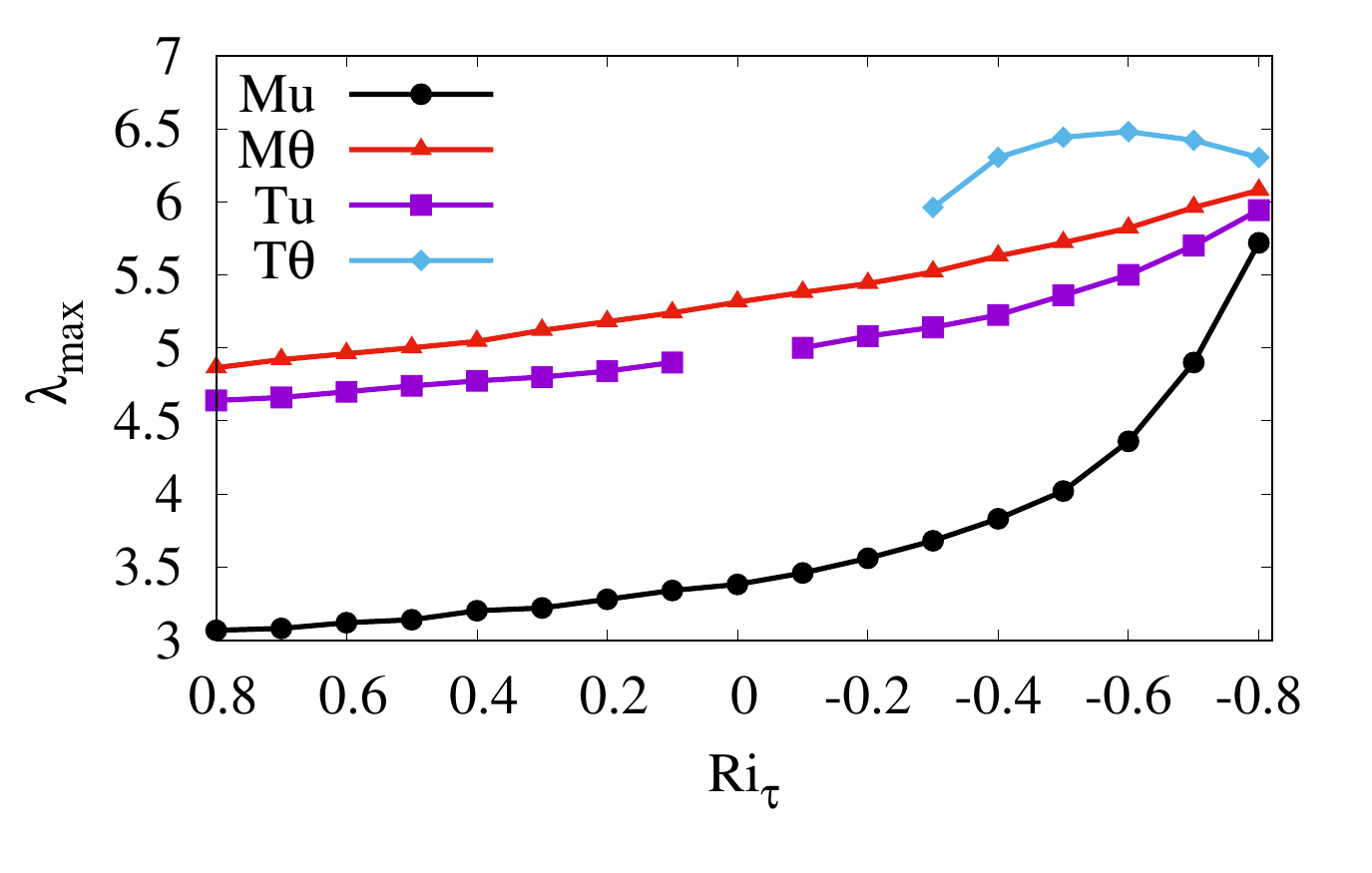}
 \put(-215,138){{$(a)$}}
 \includegraphics[width=0.44\textwidth]{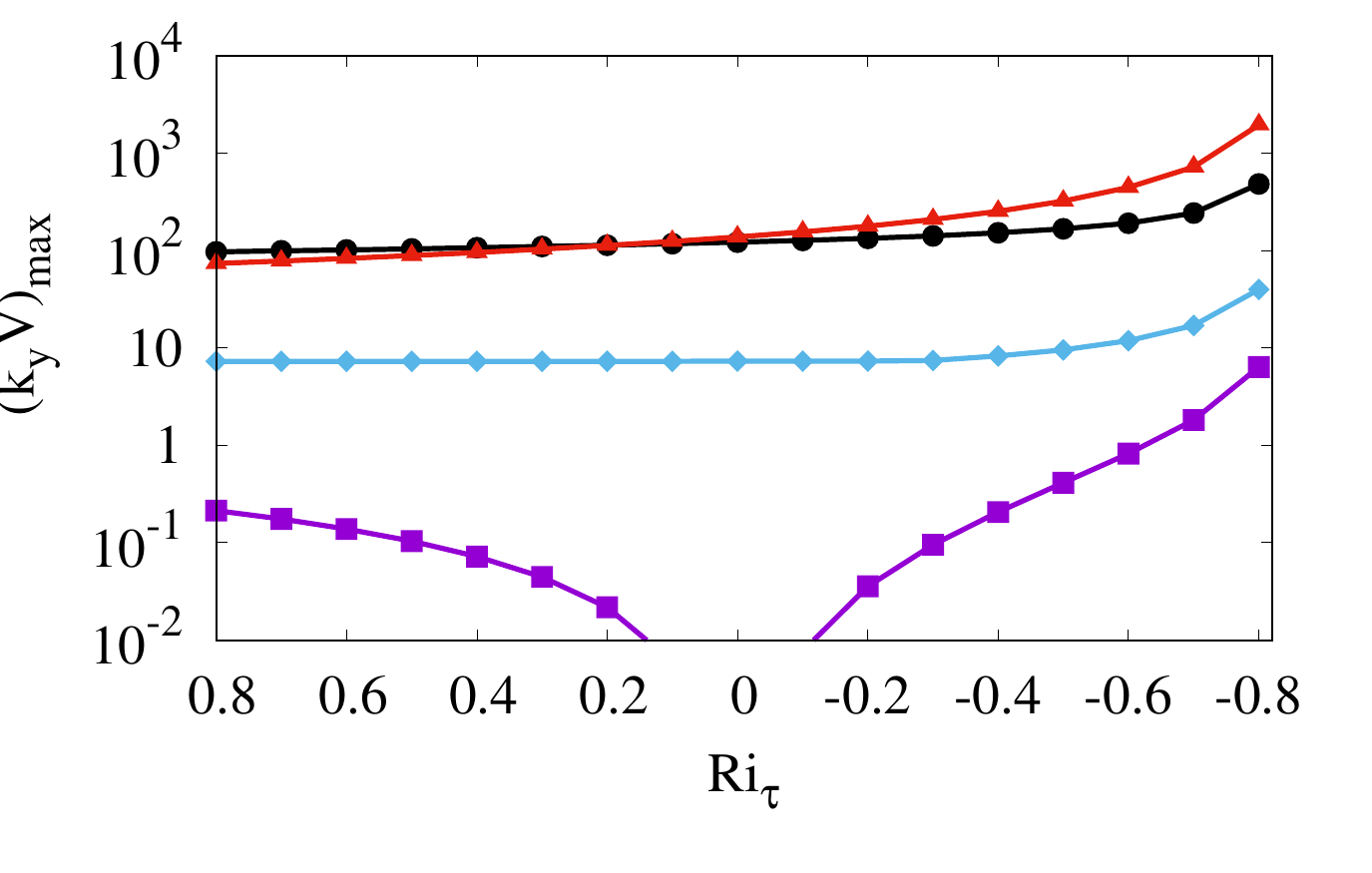} 
\put(-220,138){{$(b)$}}
}
\vspace*{-7mm}
\caption{Dependence on the Richardson number $\Ritau$ of the most amplified spanwise wavelengths $\lambda_{max}$ of streamwise-uniform modes ($k_x=0$) corresponding to the large-scale peaks (panel $a$) and of the corresponding premultiplied peak amplifications  $\Vmu$, $\Vmt$, $\Vtu$, $\Vtt$ (panel $b$) reported in the top row (panels $a$ to $d$) of \reffig{VPRElambdaChiR1000}.
  \label{fig:VPRELambdaMax}
} 
\end{figure}
The premultiplied amplifications $k_y \Vmu$ display the double-peaked structure which has already been thoroughly investigated in unstratified channels \cite{Hwang2010c}. 
The primary peak scales in outer units and corresponds to large-scale streaks with spanwise wavelengths ranging from $\approx 3$ to $\approx 6$ channel half-widths, depending on $\Ritau$. 
The secondary (lower) peak scales in wall units and corresponds to spanwise wavelengths $\lambda_y^+\approx 80-90$ ($\lambda_y=\lambda_y^+/\Retau \approx 0.085$ in outer units) typical of buffer-layer streaks \cite{Smith1983}. 
In between these two peaks is a quasi-plateau corresponding to logarithmic-layer quasi-self-similar structures \cite{delAlamo2006,Pujals2009,Cossu2009,Hwang2010c} scaling with the distance from the wall whose (non-premultiplied) variance amplification scales as $k_y^{-1}$   \cite{Hwang2010c}.
We find that the mean flow stratification has non-negligible effects only on the primary (large-scale) peak of $k_y \Vmu$ and only when $\lambda_x \gg \lambda_y$ (see Appendix B).
Indeed \reffig{VPRElambdaChiR1000}$a$ shows that for streamwise-uniform perturbations an increasingly destabilizing (stabilizing) stratification, corresponding to increasingly negative (positive) values of $\Ritau$, induces an increasing (decreasing) height of the primary peak of $k_y \Vmu$ but has no influence on  log-layer or buffer-layer structures  which have smaller spanwise wavelengths.
For perturbations with $\lambda_x= 2 \lambda_y$ stratification has no significant effect on $\Vmu$ even for large-scale structures (see \reffig{VPRElambdaChiR1000}$e$).
Similar results are found for the premultiplied temperature variance $k_y \Vmt$ produced by momentum forcing, (see \reffig{VPRElambdaChiR1000}$b$,$e$). 

The shapes of the amplification curves associated to thermal forcing differ from those associated to momentum forcing.
The premultiplied amplification curves $k_y \Vtt$, indeed, do not generally display a double-peaked shape and large-scale (large $\lambda_y$) structures are generally weakly amplified. 
For streamwise-uniform perturbations, however, unstable stratification induces the emergence of a large-scale peak while stable stratification has the opposite effect (see \reffig{VPRElambdaChiR1000}$d$) and no significant effect is found for perturbations with $\lambda_x = 2 \lambda_y$ (see \reffig{VPRElambdaChiR1000}$g$) 
Amplifications $k_y \Vtu$ are small with a single peak corresponding to large-scale structures (see \reffig{VPRElambdaChiR1000}$c$,$g$) and tend to zero $\Ritau \rightarrow 0$, as expected. 

In \reffig{VPRELambdaMax}, the most amplified spanwise wavelengths $\lambda_{max}$ and the corresponding peak of the premultiplied variance amplifications are reported as a function of $\Ritau$.
The figure shows that the spanwise wavelength maximizing $k_y V_{M\bfu}$ increases from $\lambda_y \approx 3$ (three channel half-widths) for the stably stratified case with $\Ritau=0.8$, to $\lambda_y \approx 3.5$ in the unstratified case (already examined in \refcite{Hwang2010c}) to then rapidly increase in the unstably stratified case up to $\lambda_y \approx 6$ for $\Ritau=-0.8$, near the onset of the modal instability at  $\Ri_{\tau,c}=-0.86$ (see \reffig{VPRELambdaMax}$a$).
Spanwise wavelengths maximizing $V_{M\theta}$ are larger, ranging from $\lambda_y \approx 5$ for $\Ritau=0.8$ to $\lambda_y \approx 6$ for $\Ritau=-0.8$. 
Despite this difference in the most amplified spanwise wavelengths, the peak variance of both responses to momentum forcing is very similar (see \reffig{VPRELambdaMax}$b$) being not very sensitive to $\Ritau$, except near the instability threshold for sufficiently negative $\Ritau$. 
Thermal forcing is generally less amplified than momentum forcing, particularly so for $\Vtu$ (see \reffig{VPRELambdaMax}$b$). 
The most amplified wavelengths of $\Vtu$ and $\Vmt$ are very similar while those pertaining to $\Vtt$ are, when the primary peak has emerged, larger than all other wavelengths ($\lambda_y \approx 6.5$).
The four types of variance amplification all increase when approaching the linear instability threshold $\Ritau=-0.86$, where they diverge with the most amplified wavelengths of all forced responses converging to the wavelength of the critical mode.

\subsection{Structure of the most amplified stochastic responses
\label{sec:Modes}}

\begin{figure}
\centerline{
 \includegraphics[width=0.24\textwidth]{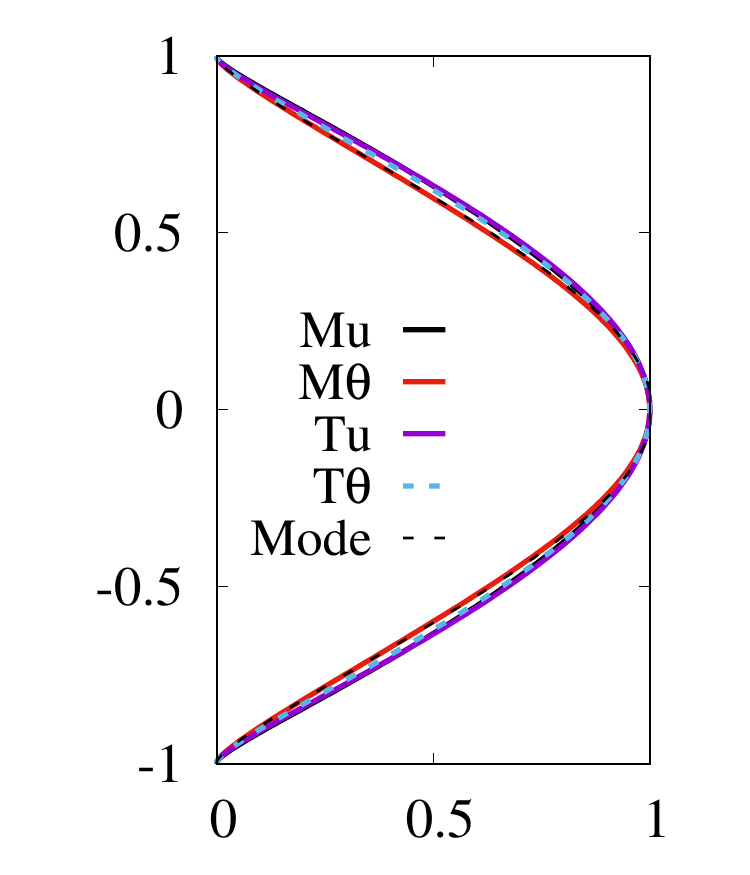} \hspace{-9mm}
 \put(-90,135){{$(a)$}}
 \put(-85,80){{$z$}}
 \put(-35,-2){{$\langle \widehat \theta \widehat w \rangle$}}
 \includegraphics[width=0.24\textwidth]{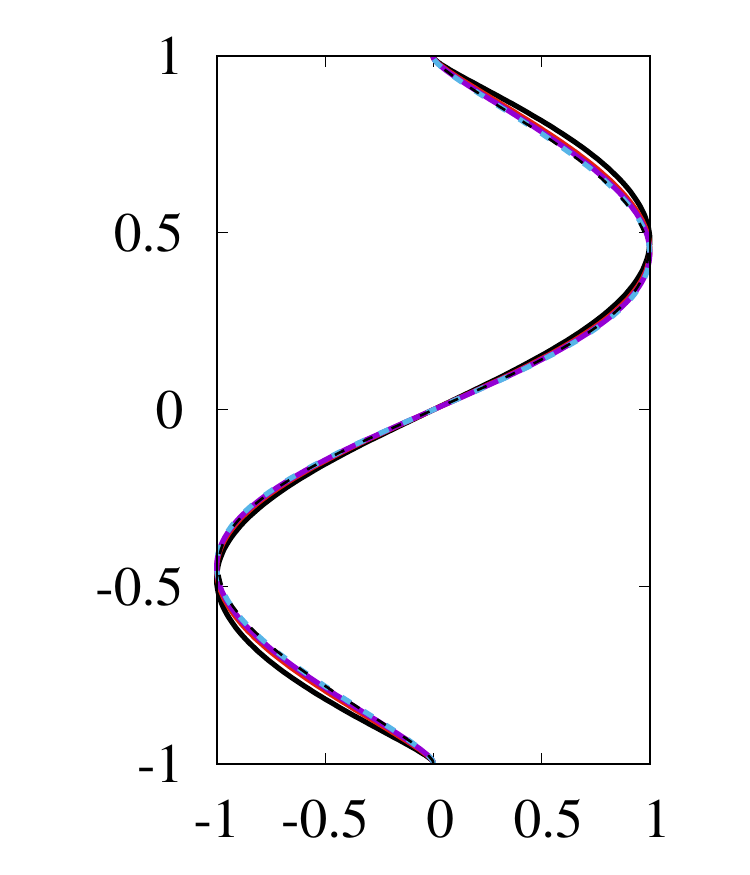} \hspace{-9mm}
 \put(-87,135){{$(b)$}}
 \put(-85,80){{$z$}}
 \put(-35,-2){{$\langle \widehat u \widehat w \rangle$}}
 \includegraphics[width=0.24\textwidth]{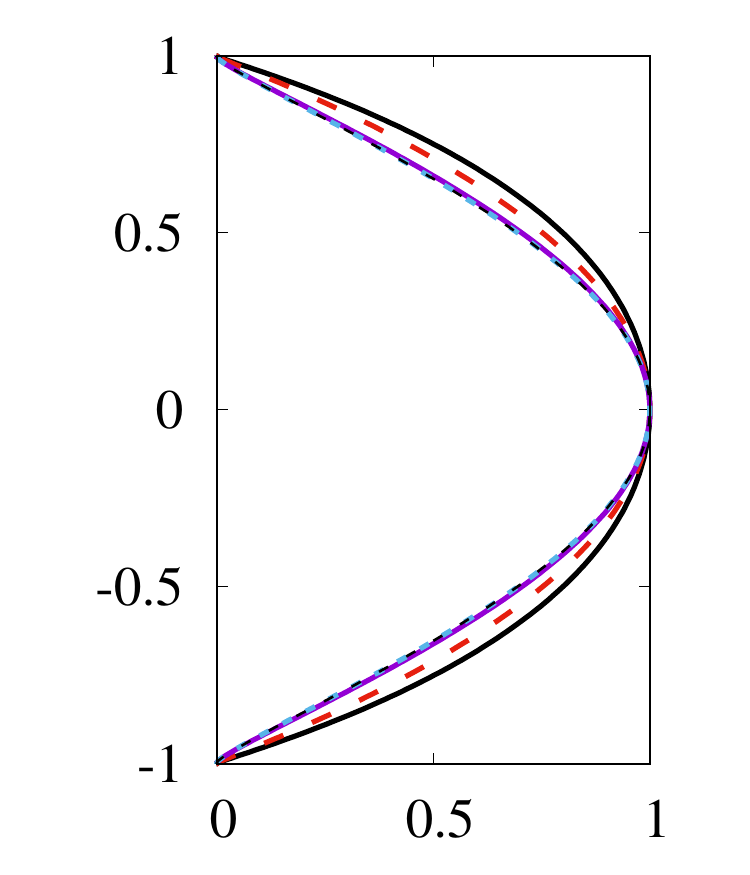} \hspace{-9mm}
 \put(-87,135){{$(c)$}}
 \put(-85,80){{$z$}}
 \put(-40,-2){{$\langle \widehat w \widehat w \rangle^{1/2}$}}
 \includegraphics[width=0.24\textwidth]{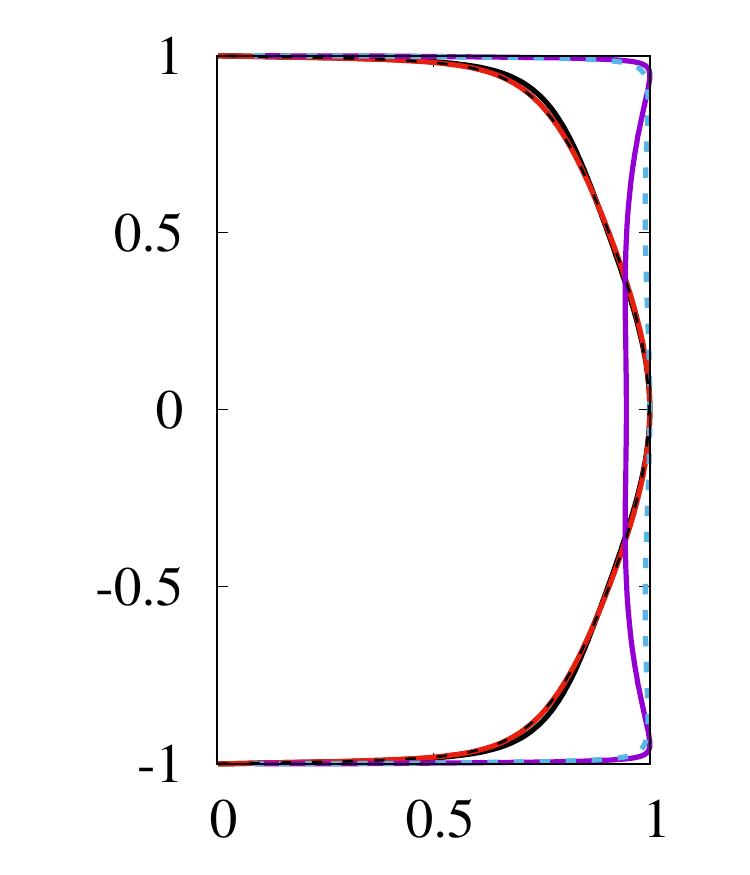} \hspace{-9mm}
 \put(-87,135){{$(d)$}}
 \put(-85,80){{$z$}}
 \put(-38,-2){{$\langle \widehat \theta \widehat \theta \rangle^{1/2}$}}
 \includegraphics[width=0.24\textwidth]{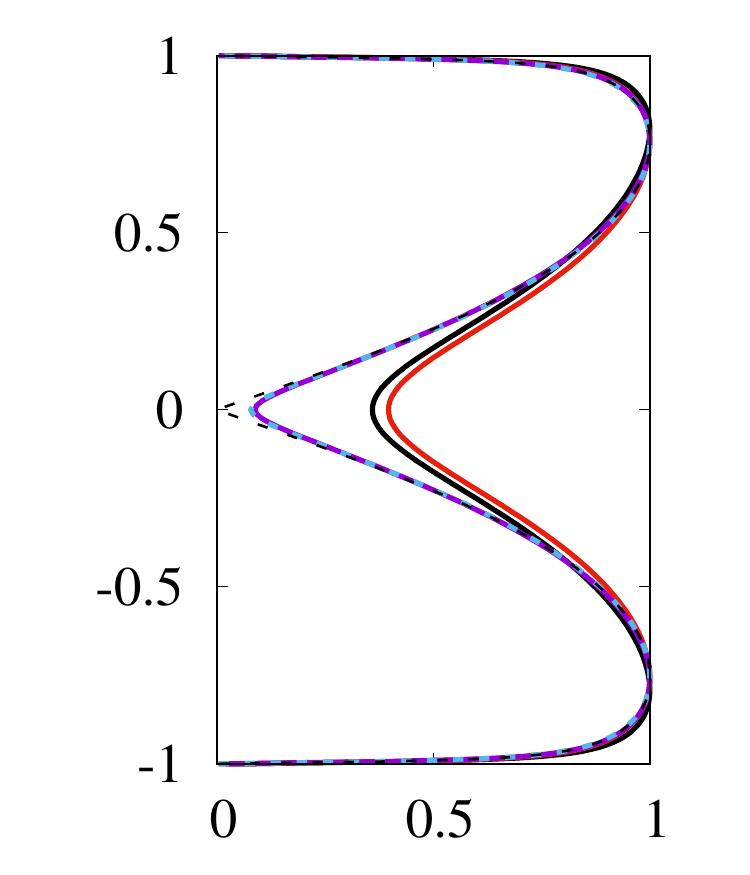} 
 \put(-110,135){{$(e)$}}
 \put(-105,80){{$z$}}
 \put(-60,-2){{$\langle \widehat u \widehat u \rangle^{1/2}$}}
 \vspace{-2mm}
}
  \caption{ Vertical profiles of the: 
(a) turbulent buoyancy flux $\langle \widehat \theta \widehat w \rangle$, 
(b) turbulent momentum flux  $\langle \widehat u \widehat w \rangle$, 
(c) $rms$ vertical velocity fluctuations $\langle \widehat w \widehat w \rangle^{1/2}$, 
(d) $rms$ temperature fluctuations $\langle \widehat \theta \widehat \theta \rangle^{1/2}$, 
(e) $rms$ streamwise velocity fluctuations $\langle \widehat u \widehat u \rangle^{1/2}$.
The different profiles have been computed at $\Retau=1000$ and $\Ritau=-0.4$ in correspondence to the four different spanwise wavelengths corresponding to the four types of large-scale peaks of the premultiplied variances of streamwise-uniform perturbations documented in panels $a-d$ of \reffig{VPRElambdaChiR1000}.
The profiles of the critical mode (incipient linear instability) computed in Ref.~\cite{Cossu2022} at $\Retau=1000$ and $\Ritau=-0.86$ (dashed black lines) are also reported for comparison. 
All profiles are normalized to unitary maximum amplitude.
  \label{fig:rmsprofs}
} 
\end{figure}
\begin{figure}
\vspace*{2mm}
\centerline{
\includegraphics[width=0.45\textwidth]{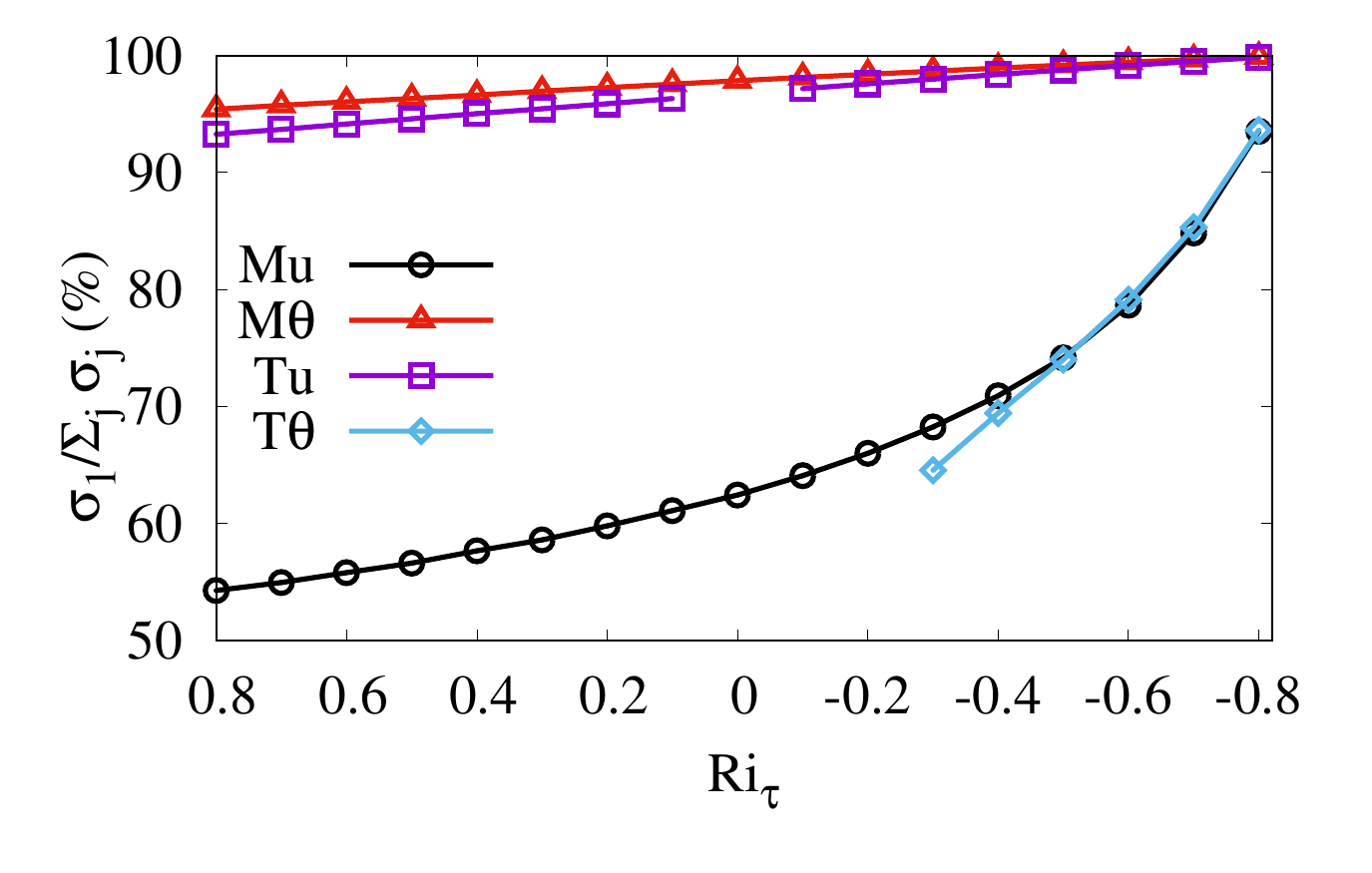} 
\put(-225,138){{$(a)$}}
\includegraphics[width=0.45\textwidth]{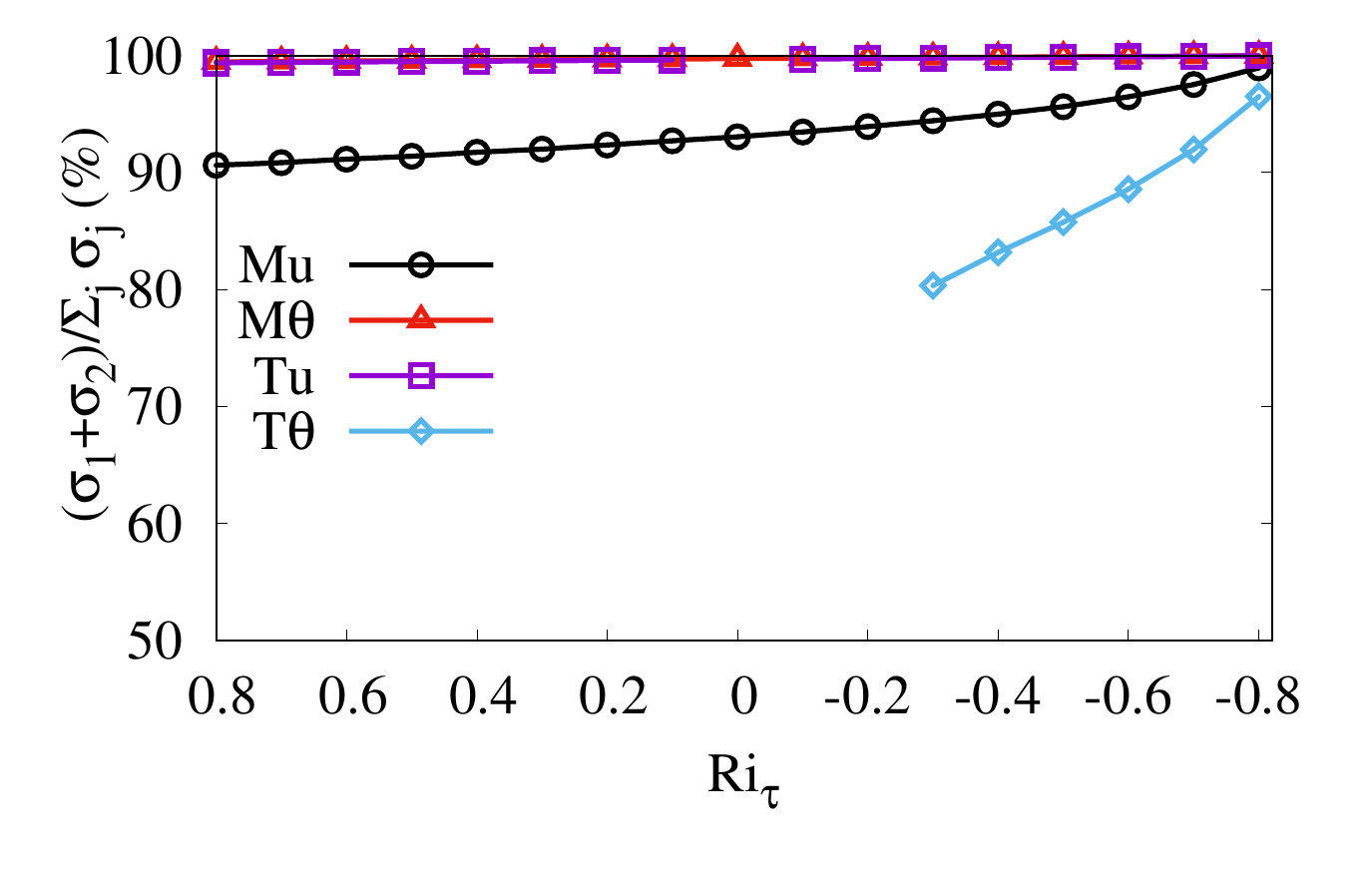}
\put(-225,138){{$(b)$}}
 \vspace{-7mm}
}
\caption{
Relative contributions of the first (panel $a$) and of the first two (panel $b$) POD modes to the peak variance of the stochastic response versus the friction Richardson number $\Ritau$ for $\Retau=1000$.}
\label{fig:KLR1000}
\end{figure}

As discussed in \refsec{Re1000}, the spanwise wavelengths maximizing the variance amplifications all converge to the spanwise wavelength of the critical mode for Richardson numbers approaching the critical value $\Ri_{\tau,c}=-0.86$.
In this limit, it is expected that the structure of the stochastic responses embed a strong signature of the critical mode.
It is, however, unclear how strong this signature is for intermediate levels of unstable stratification where important differences appear in the values of the different most amplified spanwise wavelengths (see \reffig{VPRELambdaMax}$a$) suggesting that significant differences might exist in the mechanisms underlying the different amplifications. 
We clarify this issue by examining the structure of the most amplified stochastic responses.

In \reffig{rmsprofs} are reported the vertical profiles of root mean square ($rms$) response components as well of the associated turbulent heat and momentum fluxes computed from the solutions of the Lyapunov equation, as explained in \refsec{StoFo}, for streamwise-uniform ($k_x=0$) structures having optimal spanwise wavelengths (the ones corresponding to the peak values in panels $a$-$d$ of \reffig{VPRElambdaChiR1000}) at $\Retau=1000$ 
and $\Ritau=-0.4$.
These profiles, that would otherwise have different amplitudes,  are normalized to the same (unitary) maximum amplitude in order to compare their shapes.
These profiles are also compared to those of the critical mode computed at $\Ri_{\tau,c}=-0.86$ in Ref.~\cite{Cossu2022}. 
Figs.~\ref{fig:rmsprofs}$a$ and $b$ show that the $\langle \widehat \theta \widehat w \rangle$ and $\langle \widehat u \widehat w \rangle$ profiles associated to the different large-scale peak responses are extremely similar and are also extremely similar to those of the critical mode despite their very different spanwise wavelengths (see \reffig{VPRELambdaMax}$a$) and the different $\Ritau$ for the critical mode.
This similarity reveals that, almost unexpectedly, a common vertical buoyancy and momentum transport mechanism underlies the onset of the linear instability and the amplification of stochastic forcing even for relatively weak levels of unstable stratification. 
This is further confirmed by the strong similarity of the $\langle \widehat w \widehat w \rangle^{1/2}$ profiles (see \reffig{rmsprofs}$c$) of the  $rms$ vertical velocity which is the key ingredient of the vertical turbulent transport.
Differences, however, appear between the vertical profiles of the $rms$ temperature and streamwise velocity perturbations.
A direct and `unfiltered' signature of the forcing is, indeed, clearly visible on the temperature response to thermal forcing (see \reffig{rmsprofs}$d$) that is almost-constant in the bulk of the flow because not immediately filtered by the non-normal couplings of the system.
In the case of the streamwise velocity,  differences in the response $rms$ profiles are confined to the central part of the channel (see \reffig{rmsprofs}$e$), where the mean shear is small the coupling to the vertical velocity negligible.

These results suggest that the response to stochastic forcing is composed of (a) an endogenous dominant (most amplified) highly correlated part with structure similar to that of the critical mode and (b) a more direct response to the forcing which is much less amplified/filtered and which, similarly to the forcing itself, lacks of cross-correlation between different response components.
To quantify the relative weight of these two different components in the stochastic response, a POD analysis is performed in correspondence to each of the considered peak responses for  Richardson numbers ranging from $\Ritau=0.8$ to $\Ritau=-0.8$ (always with $\Retau=1000$).

In \reffig{KLR1000} the relative contribution to the total variance of the leading (panel $a$) and of the two leading (panel $b$) POD modes are reported as a function of $\Ritau$.
The results confirm that the two leading POD modes associated to the mechanically-forced velocity large-scale peak contribute for respectively $\approx60$\% and $\approx30$\% of the $\Vmu$ variance in the unstratified case \refcite{Hwang2010c} .
We find that the contribution of these two leading POD modes to $\Vmu$ increases for increasingly destabilizing stratifications (increasingly negative values of $\Ritau$), heading towards $100$\% at $\Ri_{\tau,c}$.
A similar trend is observed for the contribution of the leading POD mode to the large-scale peak of the $\Vtt$, which is in the similar situation of having a direct contribution of the forcing to the observed variance, but with a less important contribution of the second POD mode.
On the contrary, for the $\Vtu$, $\Vmt$ amplifications, where the observed response is only indirectly forced, the contribution of the first POD mode is always larger than $\approx 95$\% ($\approx 100$\% for the first two modes) for all the considered values of $\Ritau$. These trends confirm that a strongly coherent large-scale mode is responsible of the emergence of the primary peak in the forcing response variances and is responsible of the observed strong similarity of the cross-correlations.

\subsection{Influence of the Reynolds number
\label{sec:ReDep}}

\begin{figure}
\centerline{
 \includegraphics[width=0.28\textwidth]{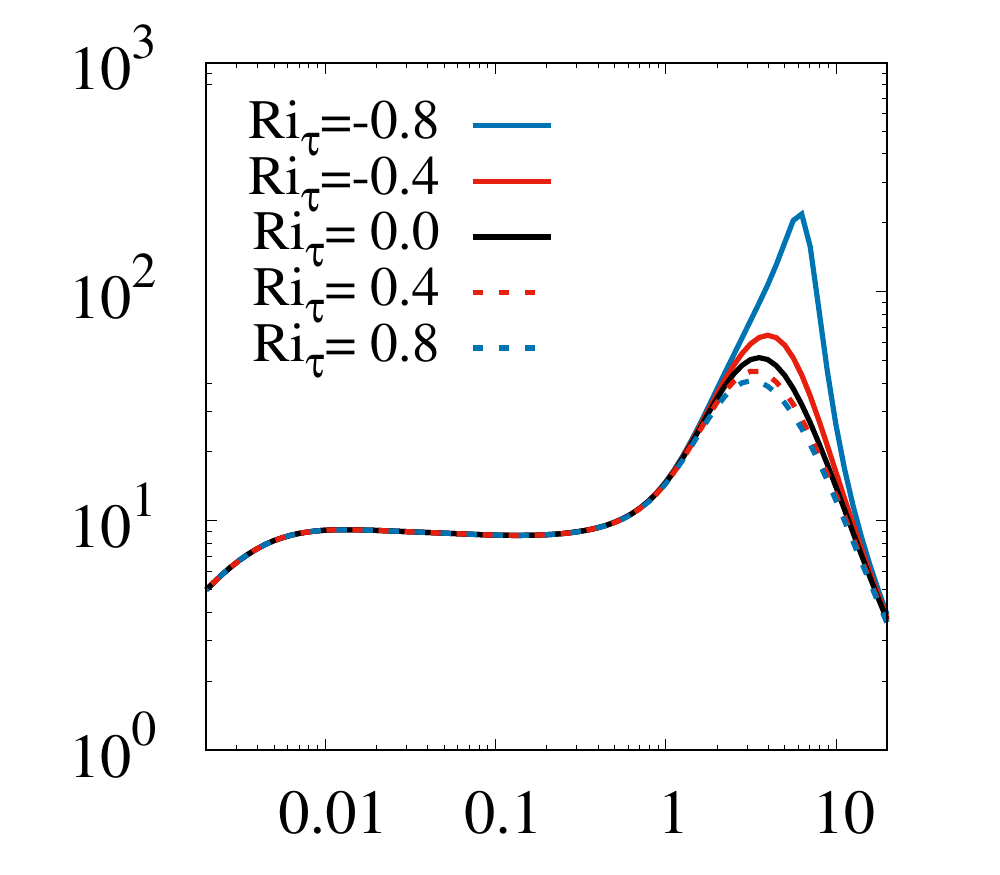} \hspace{-4mm}  
 \put(-130,120){{$(a)$}}
 \put(-130,70){{$\Vmu$}}
 \put(-52,1){{$\lambda_y$}}
 \includegraphics[width=0.28\textwidth]{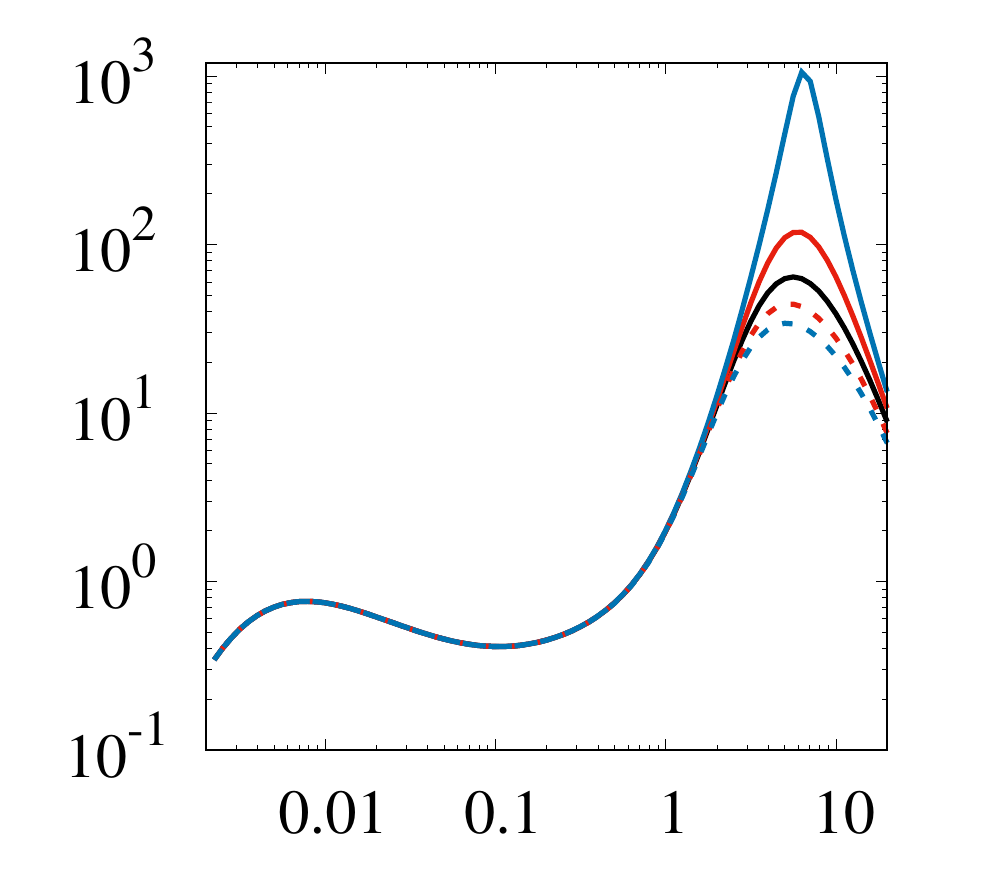} \hspace{-4mm}
 \put(-130,120){{$(b)$}}
 \put(-132,77){{$\Vmt$}}
 \put(-52,1){{$\lambda_y$}}
 \includegraphics[width=0.28\textwidth]{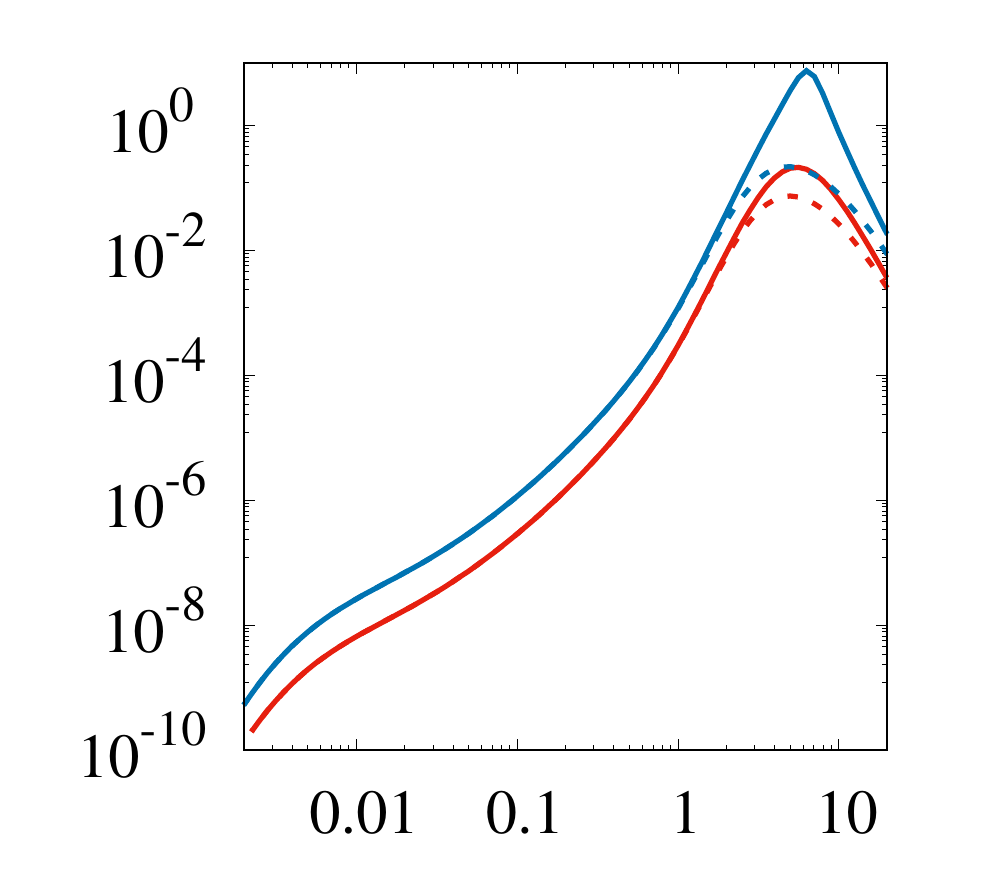} \hspace{-4mm}
 \put(-125,120){{$(c)$}}
 \put(-130,77){{$\Vtu$}}
 \put(-52,1){{$\lambda_y$}}
 \includegraphics[width=0.28\textwidth]{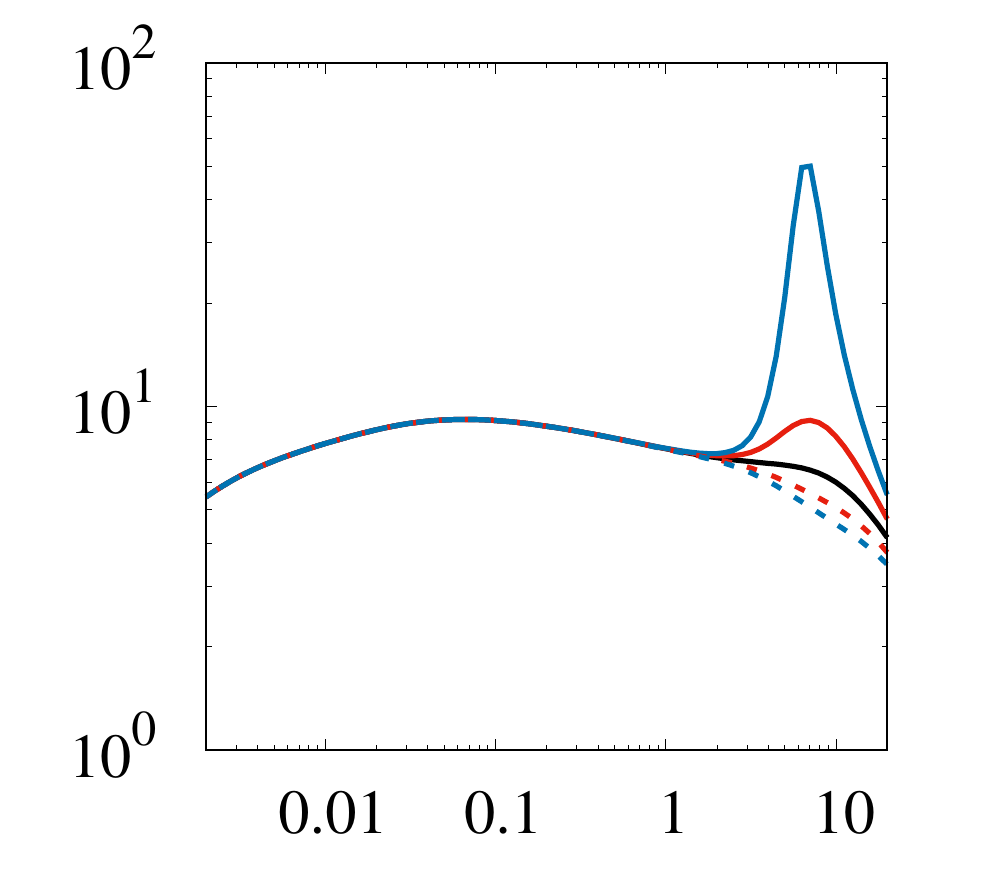} \hspace{-4mm}
 \put(-130,120){{$(d)$}}
 \put(-125,80){{$\Vtt$}}
 \put(-52,1){{$\lambda_y$}}
}
\vspace*{-3mm}
  \caption{
  Premultiplied variance amplification ratios 
$ \Vmu$   (panel $a$),
$\Vmt$ (panel $b$)
$\Vtu$   (panel $c$),
$\Vtt$ (panel $d$) of streamwise-uniform perturbations versus the spanwise wavelength $\lambda_y=2 \pi / k_y$ for 
$\Retau=10000$ and selected values of $\Ritau$. 
  \label{fig:VPRElambdaChiR10000}
} 
\end{figure}
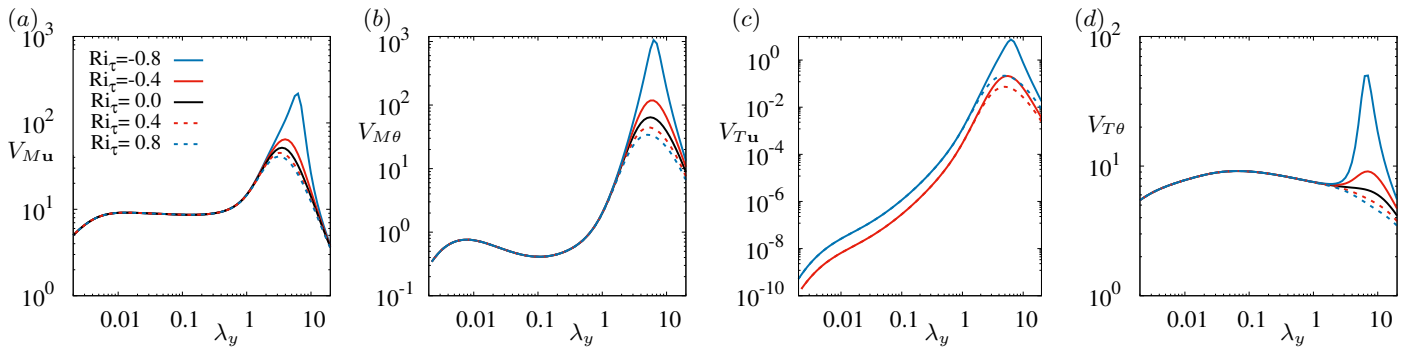

All the results discussed so far pertain to the Reynolds number $\Retau=1000$, a value typical of current direct numerical simulations (DNS) capabilities but which is on the lower end of regimes relevant to most industrial and geophysical applications.
To examine the influence of the Reynolds number on the findings discussed above, additional responses to stochastic forcing have been computed for Reynolds numbers extending from $\Retau=500$ to  $\Retau=20000$ (the highest Reynolds number considered in related previous investigations \cite{delAlamo2006,Pujals2009,Hwang2010c,Cossu2022}).

We find that for all considered Reynolds numbers, the most amplified perturbations remain the streamwise-uniform ones ($k_x=0$), which also remain the most sensitive to buoyancy effects (not shown).
Furthermore, the structure of the premultiplied variance amplification curves is found to be substantially unaffected by an increase of $\Retau$.
This can be appreciated from \reffig{VPRElambdaChiR10000} where the shown amplification curves, computed at $\Retau=10000$, remain similar to those obtained at  $\Retau=1000$ (see panels $a$-$d$ of \reffig{VPRElambdaChiR1000}), except for the increase in spatial scale separation between the primary (large-scale) and secondary (buffer-layer) peaks in the mechanically forced responses (panels $a$ and $b$ of \reffig{VPRElambdaChiR10000}).
\begin{figure}
\centerline{
\includegraphics[width=0.45\textwidth]{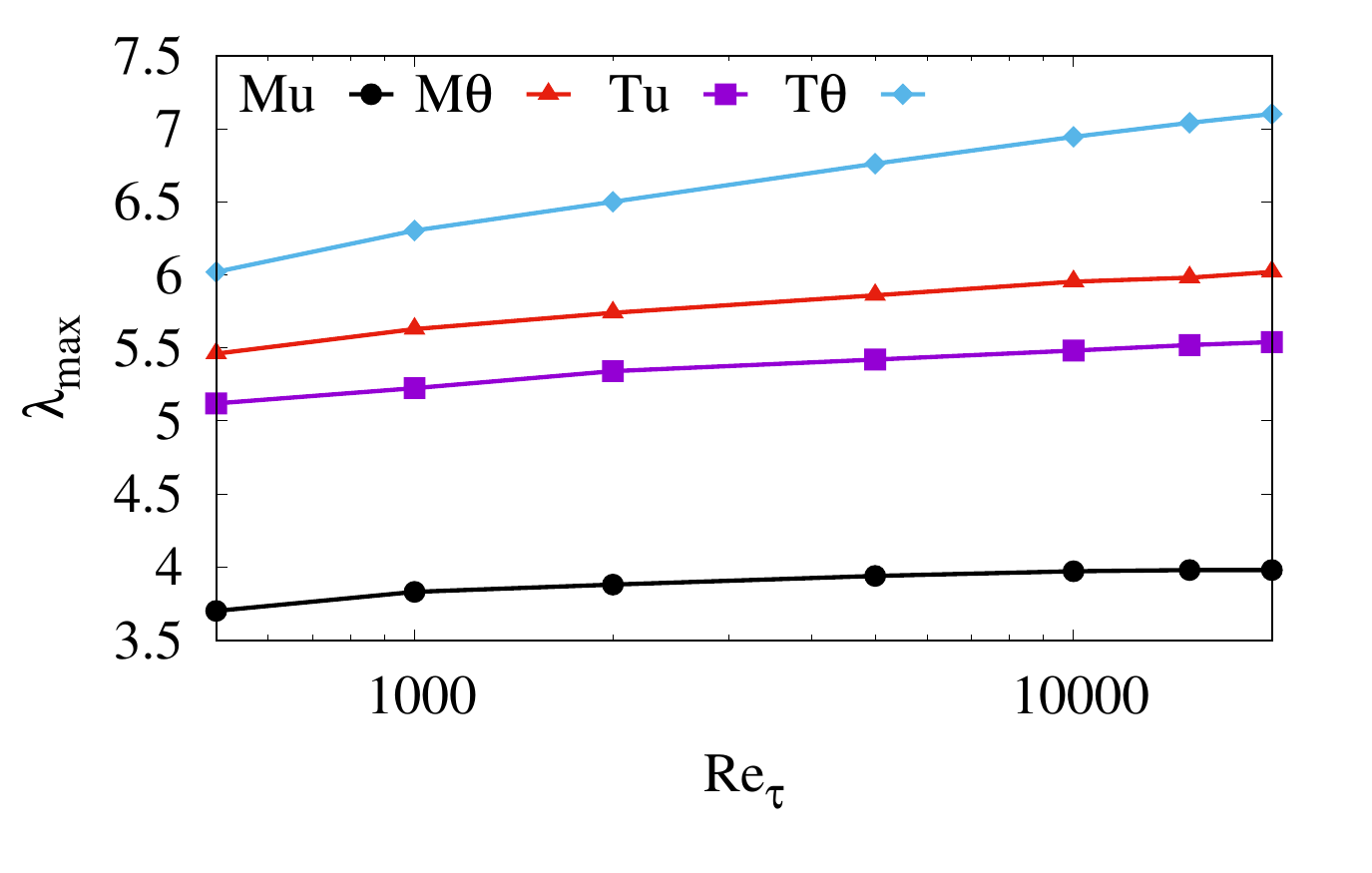} 
\put(-225,138){{$(a)$}}
\includegraphics[width=0.45\textwidth]{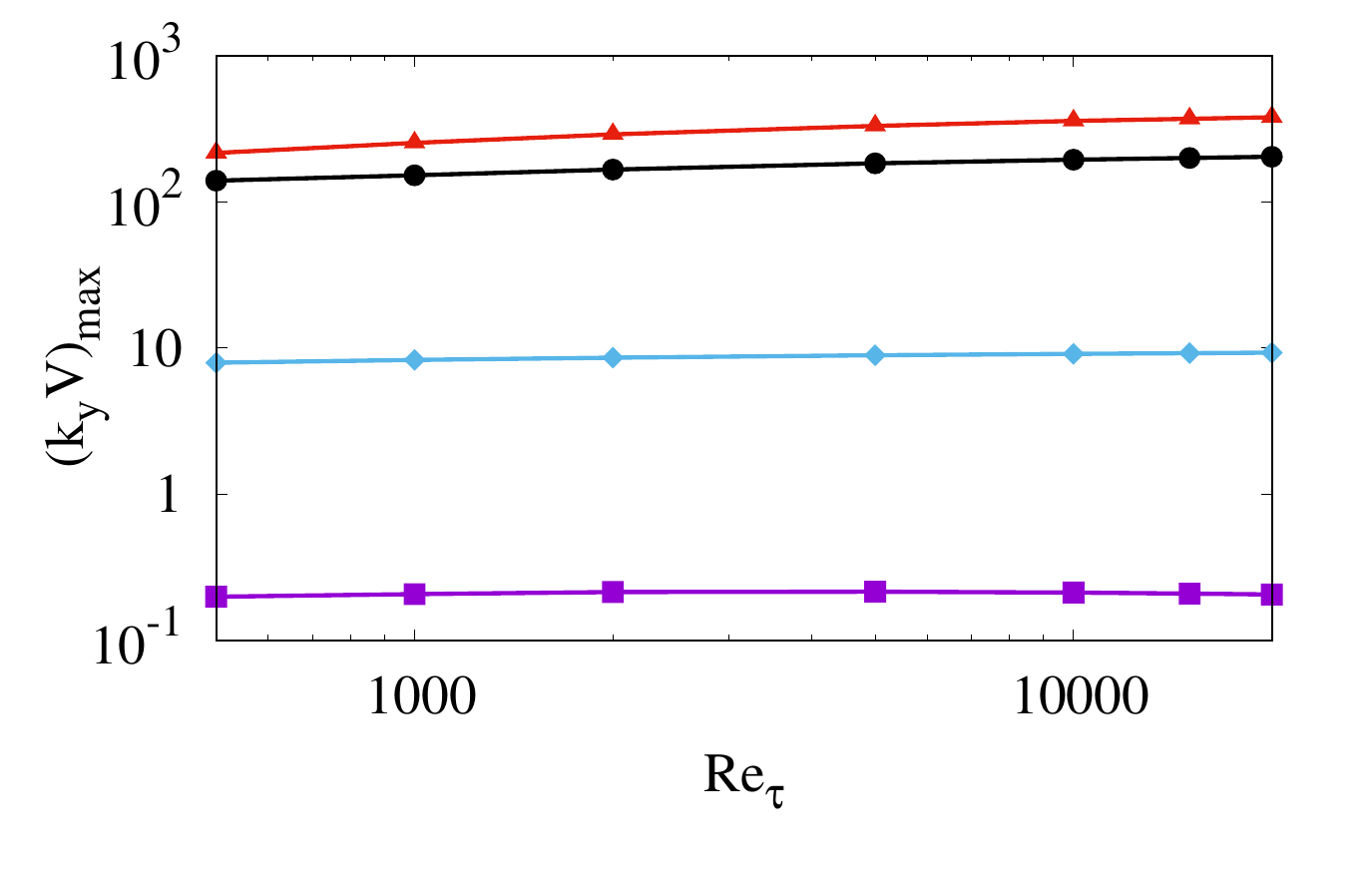} 
\put(-225,138){{$(b)$}}
}
\vspace*{-7mm}
  \caption{Reynolds number dependence of the most amplified spanwise wavelengths $\lambda_{max}$ of streamwise-uniform perturbations corresponding to the large-scale peak (panel $a$) and of the most amplified premultiplied variances (panel $b$) for the four considered types of variance amplifications $V_{Mu}$, $V_{M\theta}$, $V_{Hu}$, $V_{H\theta}$ computed for streamwise-uniform perturbations in the moderately unstably stratified regime at $\Ritau=-0.4$.
  \label{fig:VPRELambdaMaxRetau}
} 
\end{figure}
\begin{figure}
\vspace*{2mm}
\centerline{
 \includegraphics[width=0.45\textwidth]{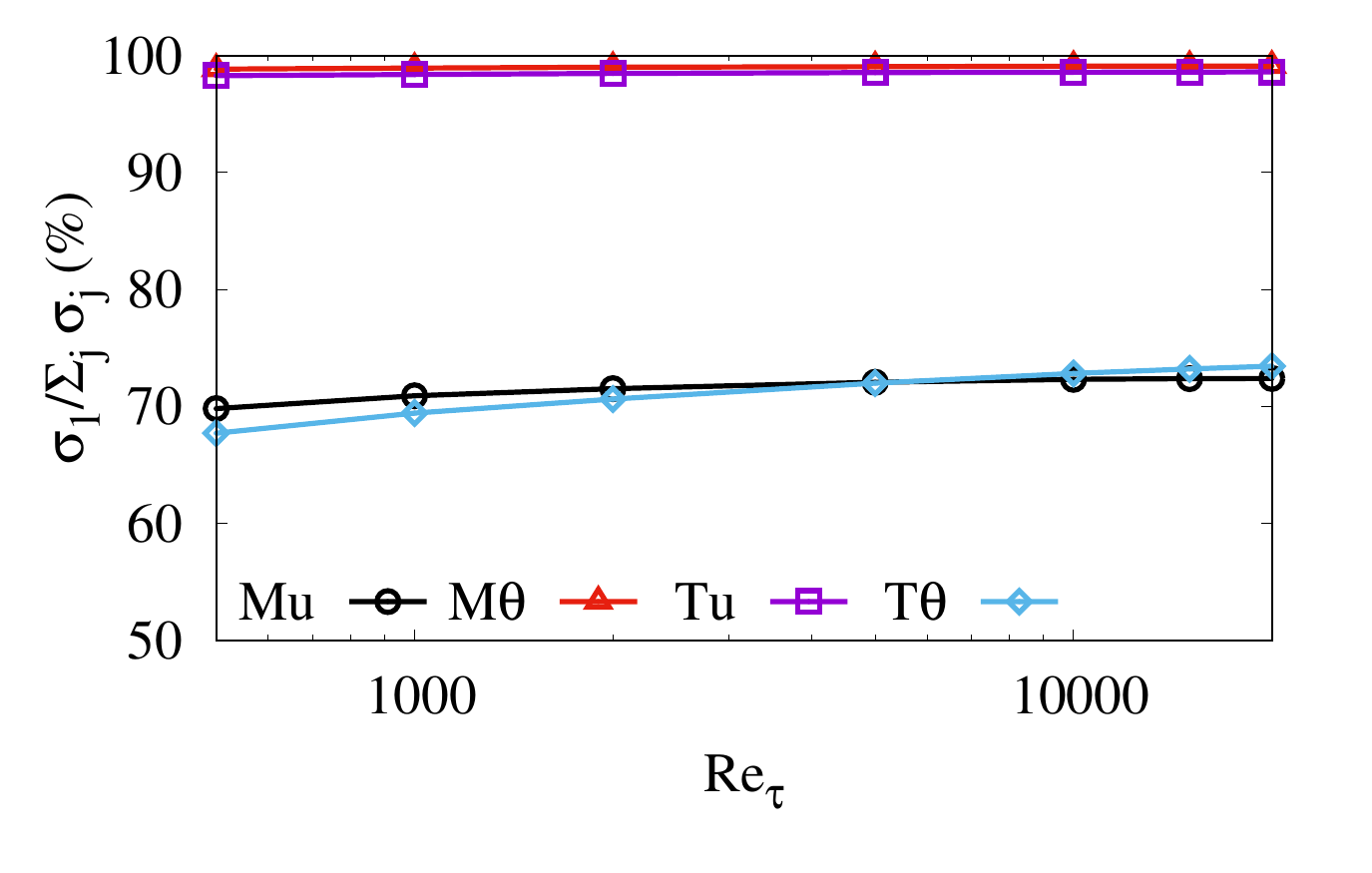} 
 \put(-225,138){{$(a)$}}
 \includegraphics[width=0.45\textwidth]{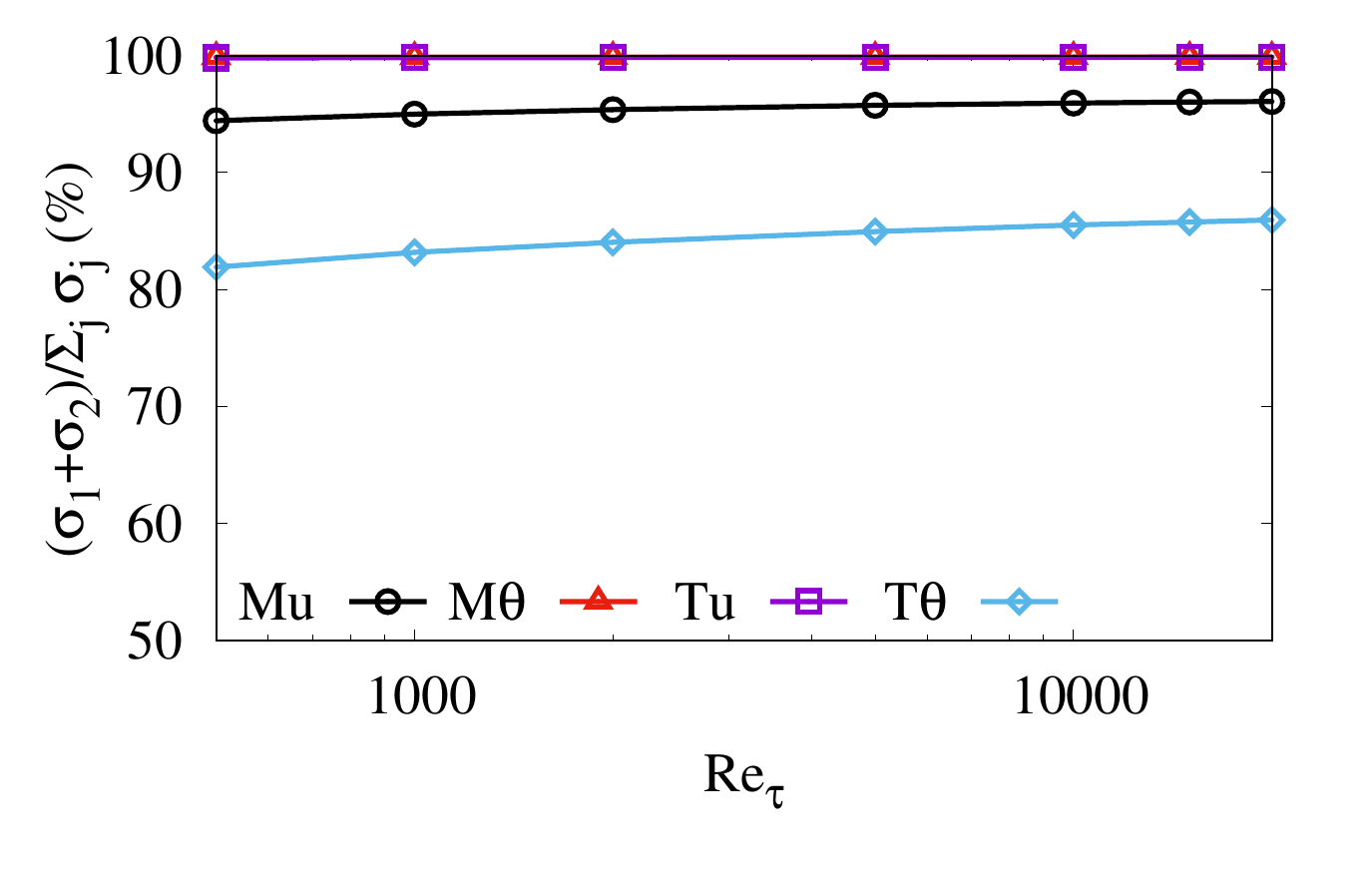}
 \put(-225,138){{$(b)$}}
 \vspace{-7mm}
}
\caption{Reynolds number dependence of the relative contributions of the first (panel $a$) and of the first two (panel $b$) POD modes to the variance of the stochastic responses computed  in correspondence to the large-scale peaks of the premultiplied variance amplifications in the moderately unstably stratified case $\Ritau=-0.4$.}
\label{fig:KLRi-040}
\end{figure}
The peak of the  (premultiplied) variance amplifications and of the associated spanwise wavelengths are reported as a function of $\Retau$ in \reffig{VPRELambdaMaxRetau} for the moderately unstably stratified case $\Ritau=-0.4$. 
This figure shows that the most amplified wavelengths increase monotonically with $\Retau$ but with only limited variations, of the order of $10$\% between $\Retau=1000$ and $\Retau=20000$. 
The maximum variance amplifications do also monotonically increase with $\Retau$, but only slightly in the considered $\Retau$ range.

Finally, we consider the influence of the Reynolds number on the relative contribution of the leading first and two POD modes to the total variance for the same moderately unstably stratified case $\Ritau=-0.4$. 
As shown in \reffig{KLRi-040}, the contributions of the first and first two POD modes to $V_{M\bfu}$ and $V_{H\theta}$ slightly increase with $\Retau$, while the leading contributions to $V_{H\bfu}$ and $V_{M\theta}$ remain $\approx 100$\%.

\section{Conclusions
\label{sec:concl}} 

The main goal of this investigation was to assess the influence of stratification on the non-normal energy amplifications of coherent perturbations in turbulent channels. 
The analysis, based on a linearized approach including the effect of turbulent Reynolds stresses in the linear operator, has been performed by computing the variance of the response to stochastic forcing.
The analysis distinguishes momentum forcing from thermal forcing and velocity fluctuations from thermal fluctuations in the response, analogously to previous investigations distinguishing different velocity and forcing components in the unstratified case \cite{Jovanovic2005}.
Results have been obtained for friction Reynolds numbers ranging from $\Retau=500$ to $\Retau=20000$, $\Pr=1$ and friction Richardson numbers ranging from $\Ritau=0.8$ (stabilizing stratification) to $\Ritau=-0.86$ (unstable stratification) corresponding to bulk Reynolds numbers and Rayleigh numbers extending up to $\Rey_b \approx 10^6$ and $\Ra \approx 10^9$.

The main findings concerning stochastic forcing amplifications are the following:
(a) momentum forcing systematically leads to larger (by at least one order of magnitude) peak variance amplifications than thermal forcing;
(b) the effect of stratification on the amplifications is non-negligible only for large-scale streamwise-elongated structures;
(c) variance amplifications do increase with increasingly unstable stratification, diverging when approaching the linear instability threshold $\Ri_{\tau,c}=-0.86$, and decrease with increasingly stable stratification, except for the variance of velocity fluctuations induced by thermal forcing, which is zero in the unstratified case;
(e) peak variance amplifications do slightly increase with the Reynolds number.

These findings are consistent with direct numerical simulations results \cite{Garcia-Villalba2011,Pirozzoli2017} showing that stable (unstable) stratification respectively lead to decreasing (increasing) thermal and velocity fluctuations but mostly at large scale and in the bulk of the flow  while leaving almost unaffected buffer-layer structures, at least when the Richardson numbers are not too large.
Our results are also reminiscent of previous findings, obtained for laminar base flows, where the influence of stratification on the optimal temporal (spatial) response to initial (boundary) conditions was found to be limited to streamwise-elongated large-scale structures \cite{Farrell1993e,Biau2004,Jerome2012}.

Concerning the spatial structure of the most amplified responses to stochastic forcing, the main findings are that:
(a) the most amplified spanwise wavelength in the velocity variance response to momentum forcing $\Vmu$, which is $\lambda_y \approx 3.5$ in the unstratified case, slightly decreases with stable stratification and increases with unstable stratification tending to $\lambda_y \approx 6$ when approaching the linear instability threshold;
(b) the most amplified spanwise wavelengths of the indirectly-forced responses (the velocity response to thermal forcing $\Vtu$ and the thermal response to momentum forcing $\Vmt$) do also increase when evolving from stable to unstable stratification, but are always larger than the $\Vmu$ most amplified spanwise wavelength;
(c) all most amplified spanwise wavelengths converge to $\lambda_y \approx 6$ when approaching the linear instability threshold;
(d) in the presence of even moderate (linearly stable) unstable stratification, all peak responses display almost indistinguishable vertical profiles of $rms$ vertical velocity fluctuations and of heat and momentum fluxes, all of which almost coincide with those of the critical mode becoming unstable at the critical Richardson number;
(e) the two most energetic POD modes of the peak responses 
contain more than 90\% of the $V_{M\bfu}$, $V_{M\theta}$ and  $V_{H\bfu}$ variance of the response.

These findings reveal that a single robust mechanism underlies the amplification of coherent large-scale structures at subcritical Richardson numbers and the onset of the instability of coherent large-scale rolls at the critical Richardson number. 
The process leading to the onset of the linear instability is gradual and the increasing response variance associated to increasingly unstable mean flow stratification as well as the increase of the optimal spanwise wavelength of the most amplified mechanically forced streaks are precursors of the linear instability of large-scale rolls. 
This suggests that at the onset of the linear instability the large-scale (subcritical) nonlinear self-sustained structures implying large-scale coherent streaks and quasi-streamwise vortices \cite{Hwang2010b,Rawat2015,Hwang2016,Cossu2017} have probably morphed into convection-driven saturated coherent rolls.
It is therefore likely that the linear analysis developed in the present investigation and in \cite{Cossu2022} can be extended into the supercritical regime by merging now `classical' methods, previously used to investigate nonlinear laminar convection \cite{Clever1977,Clever1991,Clever1992,Clever1997,Waleffe2015}, with the more recent techniques used to isolate large-scale self-sustained processes in turbulent unstratified flows \cite{Hwang2010b,Rawat2015,Rawat2016,Hwang2016}.
Such an extension to the nonlinear domain would 
probably allow for the elucidation of the nature of the transition from streamwise rolls to open cells and hopefully provide more insights into the free-convection regime.
It would be also be of great interest to apply the approach taken in this study and in \cite{Cossu2022} to atmospheric boundary layers by including a number of additional effects in the analysis such as surface roughness, the three-dimensionality of the mean velocity profile including an inflection point and the effect of Coriolis acceleration. 
These extensions are the object of current intensive effort.
In this context, however, it would be desirable to obtain, by DNS or experimentally, additional validations of our theoretical approach
at Reynolds numbers much higher than those currently available.

\appendix

\section{Turbulent mean flow model
\label{app:ExtCess}}

The temporally-averaged mean flow profiles, that are used in this study are based on the model introduced in \refcite{Cossu2022} which extends Cess's 1958 model \cite{Cess1958} to the (weakly) stratified regime. 
Such a model is adapted to the high $\Retau$ small $|\Ritau|$ regime, where the flow is fully turbulent but stratification effects are small enough that the temperature field behaves as a passive scalar. 
In this regime the $U(z)$, $\nu_T(z)$, $\Theta(z)$ and $\alpha_T(z)$ profiles do not depend on $\Ritau$.
We briefly summarize here the main features of the model for $\Pr=1$.

For the eddy viscosity and the mean velocity profiles, Cess's expressions, as reported in \refcite{Reynolds1967}, are assumed:
\begin{equation}\label{eq:CessNu}
\nu_T = \frac{1 }{ 2 \Retau} \left\{1+\left[{\kappa\, \Retau}(1-z^2)\,\frac{1+2z^2 }{ 3}\,  \left(1-e^{-z^+ / A}\right) \right]^2\right\}^{1/2}+\frac{1}{ 2 \Retau}
;~~~~~~~~~~~
\frac{dU}{dz}=-\frac{z}{\nu_T},
\end{equation}
where the mean velocity profile is obtained from integration of $dU/dz$ and the values of the von K\'{a}rm\'{a}n constant $\kappa=0.426$ and $A=25.4$ are calibrated on DNS data obtained at $\Retau=2000$ in \refcite{Hoyas2006}.
Cess's model has been widely used in linear analyses of turbulent mean flows (see e.g. Refs.~\cite{Reynolds1967,Waleffe1993,Farrell1996b,delAlamo2006,Pujals2009,Hwang2010c,Morra2019}, among others) providing reasonable approximations of the $U(z)$ profiles, a few examples of which are shown in \reffig{MeanFlow}$a$.
Furthermore, in \refcite{Cossu2022} is was shown that the friction law $C_{f}(\Rey_b)$ computed with Cess's model agrees well with DNS data reported in \refcite{Pirozzoli2017}  and fits the Prandtl's law derived therein even at Reynolds numbers much higher than those accessed in the DNS.
The Cess model has been extended to provide a reasonable fit to the mean temperature profile:
\begin{equation} \label{eq:CessAlpha}
\alpha_T = \frac{1 }{ 2 \Pr \Retau} \left\{1+\left[{\kappa\, \Retau}(1-z^2)\,\frac{1+2z^2}{3}\,\frac{1-\chi z^2}{1-\chi}\,  \left(1-e^{-z^+ / A}\right) \right]^2\right\}^{1/2}+\frac{1}{2 \Pr \Retau}
;~~~
\frac{d\Theta}{dz} = - \frac{Q }{ \alpha_T},
\end{equation}
where the same constants $\kappa$, $A$ as in \refeq{CessNu} are used, $\chi=0.25$ (see \refcite{Cossu2022} for more details) and the mean temperature profile is obtained by the usual vertical integration.
In \refcite{Cossu2022} is was shown that: (a) the model's mean temperature profiles, a few examples of which are shown in \reffig{MeanFlow}$b$,  fit reasonably well the DNS data of \refcite{Pirozzoli2017} and (b) that the $\Nu(\Rey_b)$ curve computed by means of  \refeq{CessAlpha} fits  well DNS data reported in \refcite{Pirozzoli2017} as well as the empirical fit $\Nu = 0.0073\,\Rey_b^{0.802}$ reported in the same study, and this even at Reynolds numbers much higher than those accessed in the DNS.

\section{Influence of the streamwise wavenumber
\label{app:kxeffect}}

\begin{figure}
\centerline{
 \includegraphics[width=0.28\textwidth]{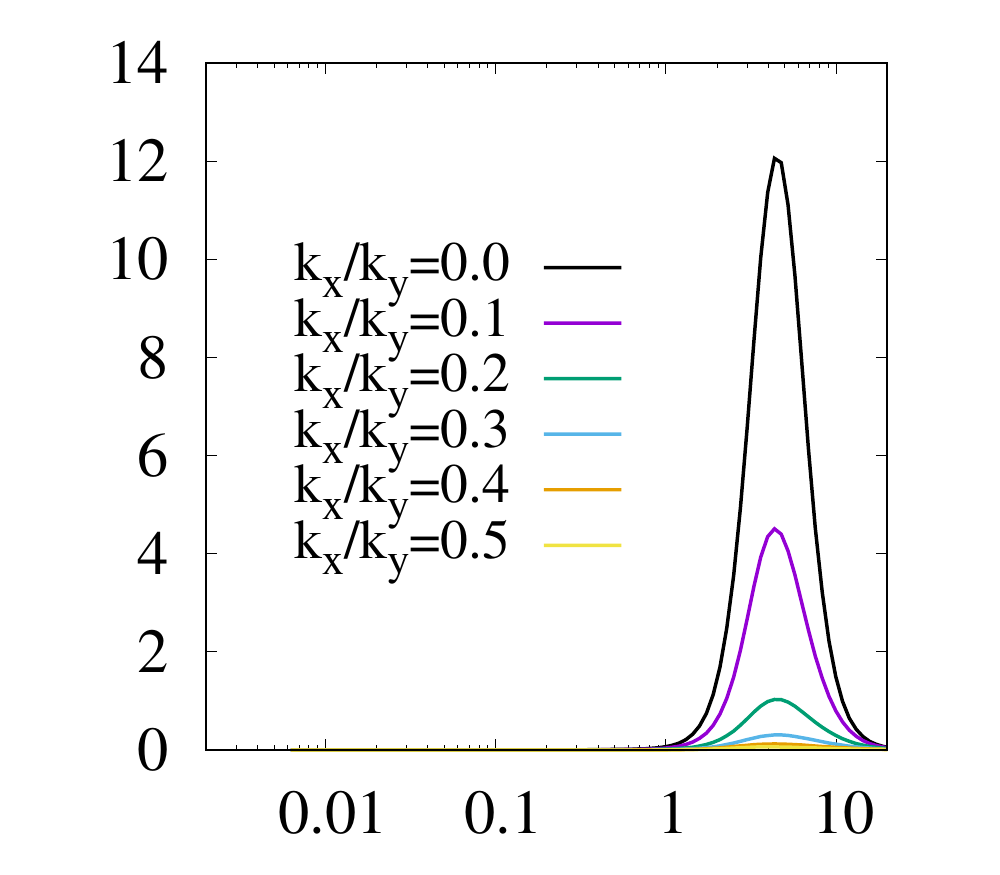} \hspace{-4mm}  
 \put(-130,120){{$(a)$}}
 \put(-80,100){{$k_y (\Delta \Vmu)$}}
 \put(-52,1){{$\lambda_y$}}
 \includegraphics[width=0.28\textwidth]{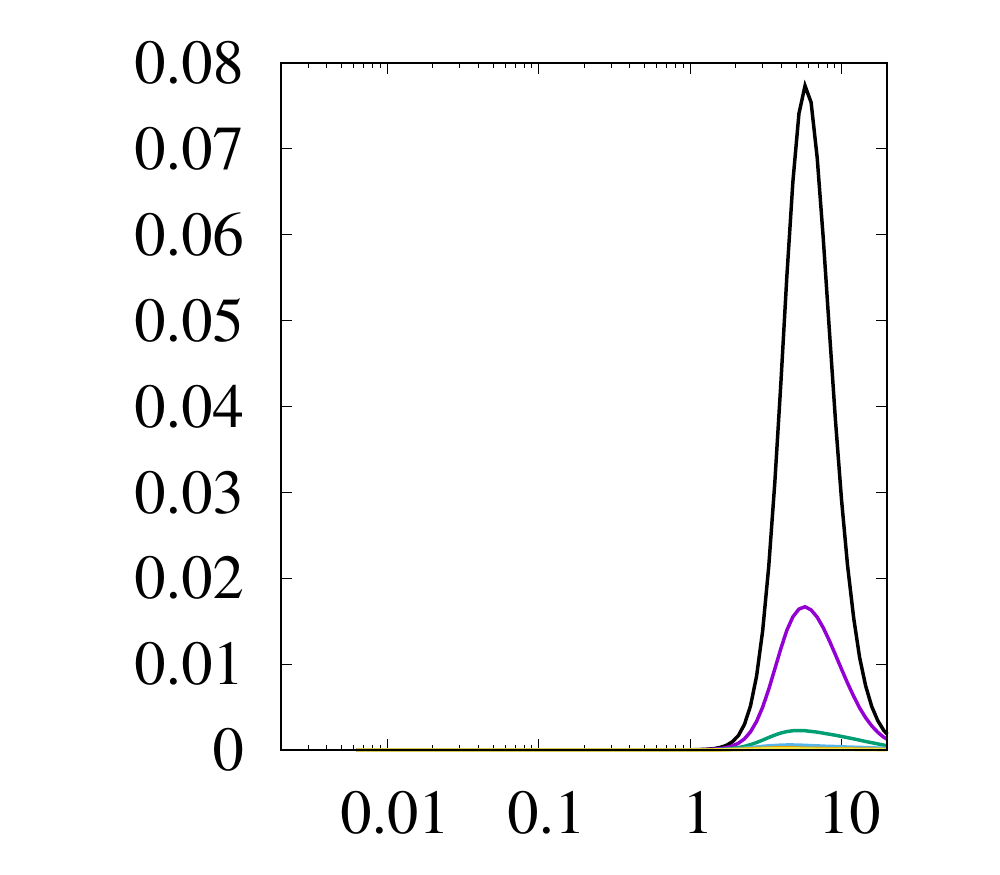} \hspace{-4mm}
 \put(-130,120){{$(b)$}}
 \put(-80,100){{$k_y (\Delta  \Vmt)$}}
 \put(-52,1){{$\lambda_y$}}
 \includegraphics[width=0.28\textwidth]{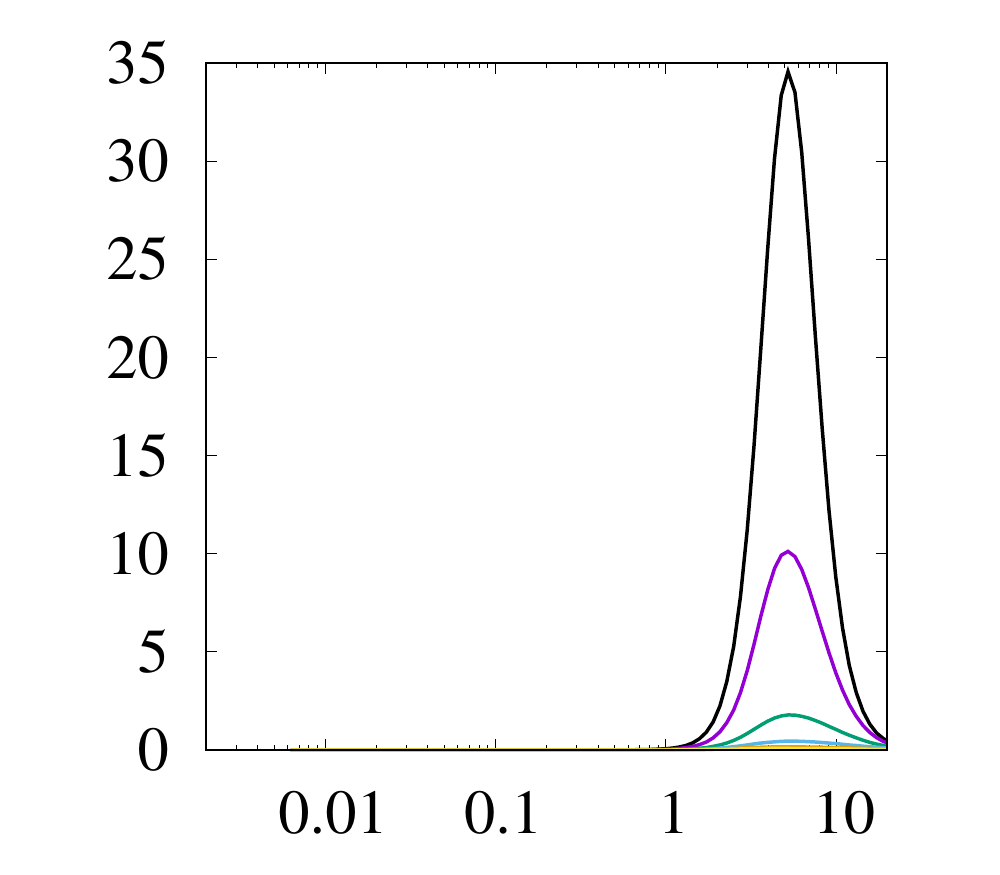} \hspace{-4mm}
 \put(-130,120){{$(c)$}}
 \put(-80,100){{$k_y (\Delta  \Vtu)$}}
 \put(-52,1){{$\lambda_y$}}
 \includegraphics[width=0.28\textwidth]{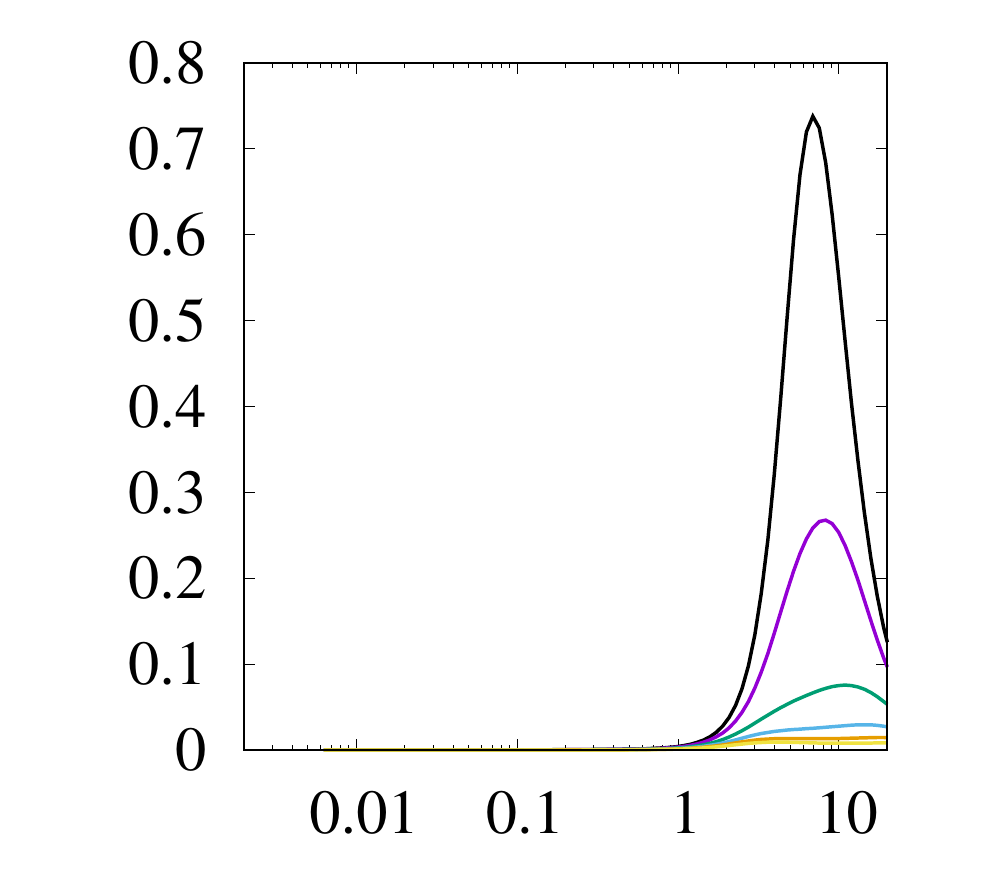} \hspace{-4mm}
 \put(-130,120){{$(d)$}}
 \put(-80,100){{$k_y (\Delta \Vtt)$}}
 \put(-52,1){{$\lambda_y$}}
}
\vspace*{-1mm}
\centerline{
 \includegraphics[width=0.28\textwidth]{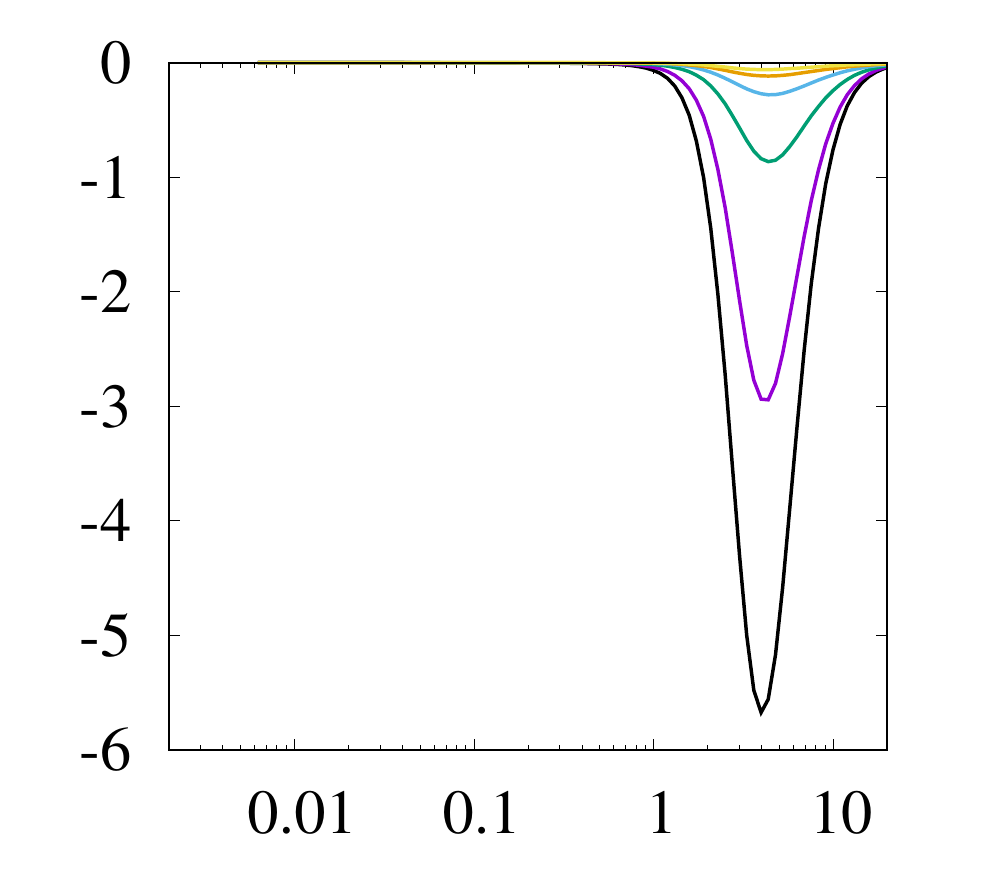} \hspace{-4mm}
 \put(-130,120){{$(e)$}}
 \put(-90,100){{$k_y (\Delta \Vmu)$}}
 \put(-52,1){{$\lambda_y$}}
 \includegraphics[width=0.28\textwidth]{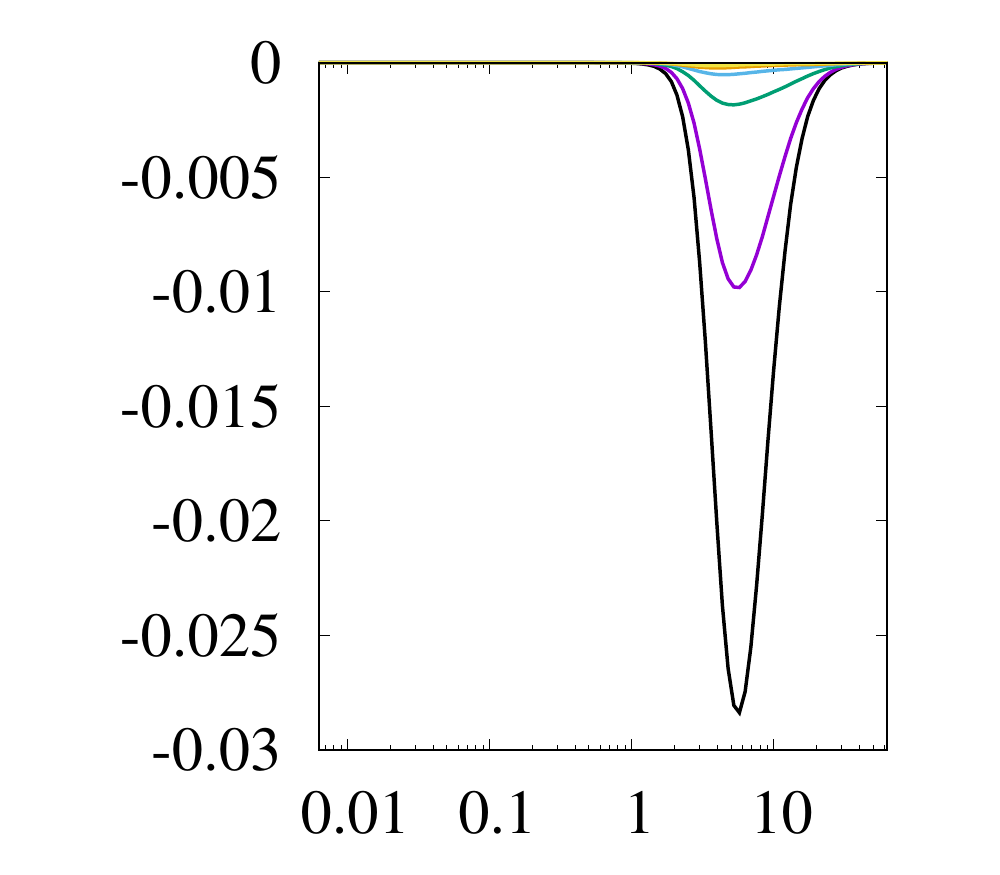} \hspace{-4mm}
 \put(-130,120){{$(f)$}}
 \put(-85,100){{$k_y (\Delta \Vmt)$}}
 \put(-52,1){{$\lambda_y$}}
 \includegraphics[width=0.28\textwidth]{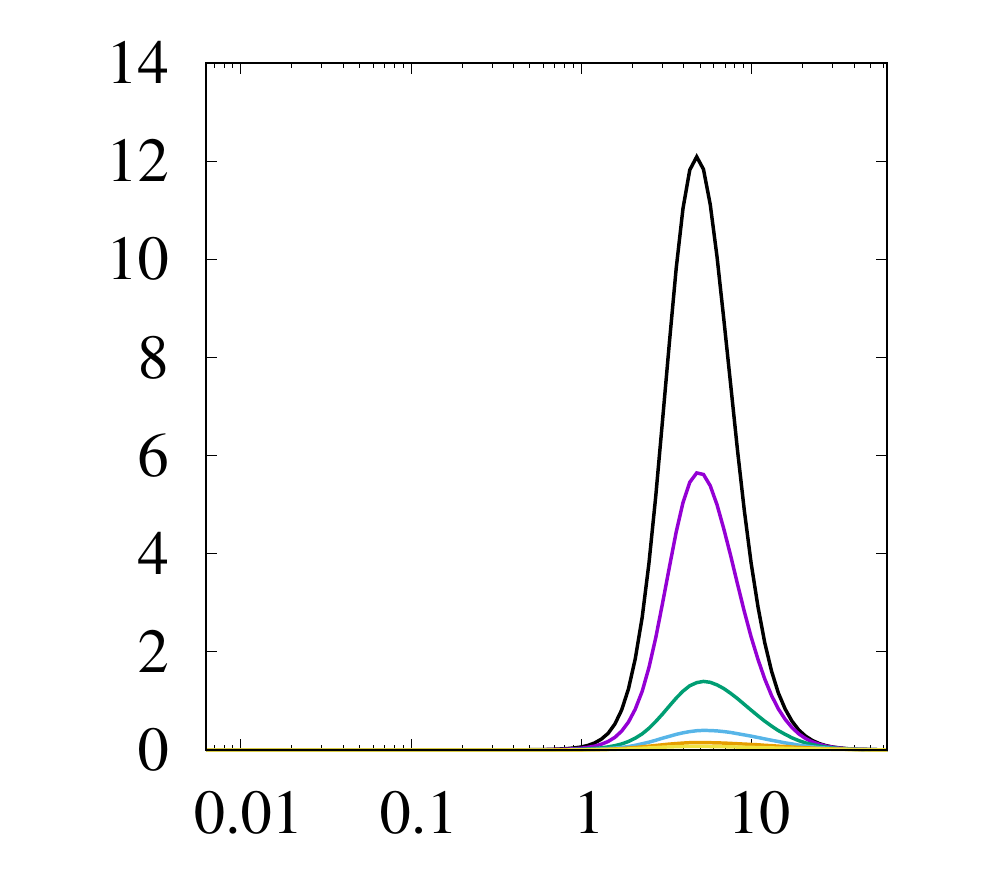} \hspace{-4mm}
 \put(-130,120){{$(g)$}}
 \put(-85,100){{$k_y (\Delta \Vtu)$}}
 \put(-52,1){{$\lambda_y$}}
 \includegraphics[width=0.28\textwidth]{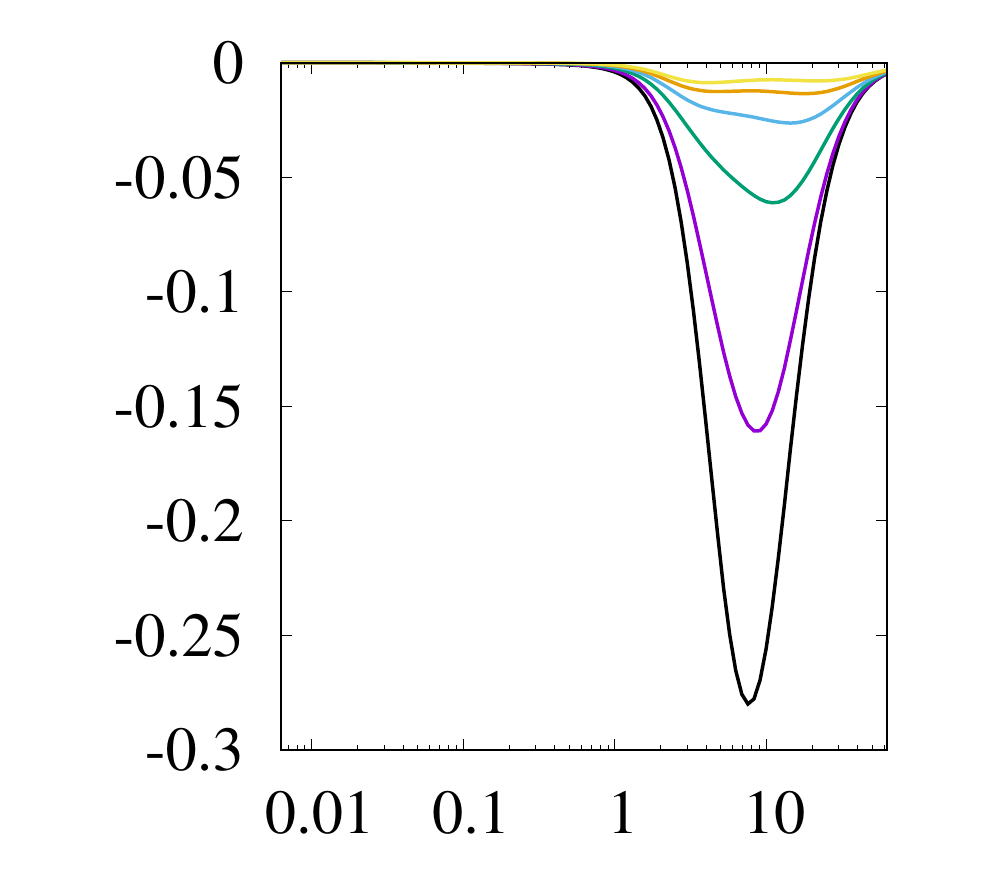} \hspace{-4mm}
 \put(-130,120){{$(h)$}}
 \put(-85,100){{$k_y (\Delta \Vtt)$}}
 \put(-52,1){{$\lambda_y$}}
}
\vspace*{-3mm}
\caption{Stratification-induced variations $\Delta V(\Ritau,k_x,k_y)= V(\Ritau,k_x,k_y)- V(0,k_x,k_y)$ of the variance amplification ratios
$\Vmu$   (panels $a$ and $e$),
$\Vmt$ (panels $b$ and $f$),
$\Vtu$   (panels $c$ and $g$) and
$\Vtt$ (panels $d$ and $h$) 
for the unstably stratified case $\Ritau=-0.4$ (top row, panels $a$ to $d$) and 
the stably stratified case $\Ritau=0.4$ (bottom row, panels $e$ to $h$) for selected ratios $k_x/k_y$ of the streamwise to spanwise wavenumber and $\Retau=1000$.
The variations are reported in premultiplied form $k_y (\Delta V)$ as a function of the spanwise wavelength $\lambda_y=2 \pi / k_y$} 
\label{fig:VdiffPRElambdaRiR1000}
\end{figure}

In \refsec{Re1000}, the effect of stratification on  variance amplifications was investigated for streamwise-uniform perturbations (i.e. for $k_x=0$) and for the case where the streamwise wavelength was twice the spanwise wavelength $\lambda_x/\lambda_y=2$ (i.e. $k_x/k_y=1/2$).
It was shown that stratification effects on the variance amplifications were negligible when $k_x/k_y=1/2$ and the analysis was therefore subsequently focused on the $k_x=0$ case.
In this appendix we report additional results for values of $k_x/k_y$ intermediate between $0$ an $0.5$ to confirm that the maximum sensitivity to stratification effects is found for streamwise-uniform perturbations ($k_x=0$) and not, e.g., for (finite) streamwise wavelengths longer that $2 \lambda_y$.
To this end, in \reffig{VdiffPRElambdaRiR1000} are reported the stratification-induced variations $\Delta V(\Ritau,k_x,k_y)= V(\Ritau,k_x,k_y)- V(0,k_x,k_y)$ of the variance amplification ratios 
$\Vmu$, $\Vmt$, $\Vtu$,$\Vtt$  with respect to the neutrally-stratified case. 
The moderately unstably-stratified case $\Ritau=-0.4$ and moderately stably-stratified case $\Ritau=0.4$ and the wavelength ratios   
$\lambda_x=10 \lambda_y$ ($k_x/k_y=0.1$),
$\lambda_x=5 \lambda_y$ ($k_x/k_y=0.2$),
$\lambda_x=3.33 \lambda_y$ ($k_x/k_y=0.3$),
$\lambda_x=2.5 \lambda_y$ ($k_x/k_y=0.4$) are considered.

From \reffig{VdiffPRElambdaRiR1000} it is seen that the maximum deviations from the neutrally-stratified case are effectively obtained for streamwise-uniform perturbations and that these deviations monotonically decrease with decreasing values of $\lambda_x/\lambda_y$ (i.e. with increasing values of $k_x/k_y$).
For all considered cases these deviations are significant only for large-scale structures with $\lambda_y=O(1-10)$.


\newcommand{\noopsort}[1]{} \newcommand{\printfirst}[2]{#1}
  \newcommand{\singleletter}[1]{#1} \newcommand{\switchargs}[2]{#2#1}
%

\end{document}